\documentclass[acmsmall]{acmart}
\usepackage{makecell}
\usepackage{subcaption}
\usepackage{xcolor}
\usepackage{multicol}
\usepackage[most]{tcolorbox}
\AtBeginDocument{%
  \providecommand\BibTeX{{%
    \normalfont B\kern-0.5em{\scshape i\kern-0.25em b}\kern-0.8em\TeX}}}

\setcopyright{acmlicensed}
\acmJournal{PACMHCI}
\acmYear{2024} \acmVolume{8} \acmNumber{CSCW1} \acmArticle{80} \acmMonth{4} \acmPrice{15.00}\acmDOI{10.1145/3637357}

\newcommand{\hlwq}[1]{{\textcolor{red}{#1}}}

\begin{document}

\title{Investigating VTubing as a Reconstruction of Streamer Self-Presentation: Identity, Performance, and Gender}

\author{Qian Wan}
\affiliation{%
  \institution{City University of Hong Kong}
  \streetaddress{83 Tat Chee Avenue}
  \city{Hong Kong}
  \country{China}
}
\email{qianwan3-c@my.cityu.edu.hk}

\author{Zhicong Lu}
\affiliation{%
  \institution{City University of Hong Kong}
  \streetaddress{83 Tat Chee Avenue}
  \city{Hong Kong}
  \country{China}
  }
\email{zhiconlu@cityu.edu.hk}



\begin{abstract}
  VTubers, or Virtual YouTubers, are live streamers who create streaming content using animated 2D or 3D virtual avatars. In recent years, there has been a significant increase in the number of VTuber creators and viewers across the globe. This practice has drawn research attention into topics such as viewers' engagement behaviors and perceptions, however, as animated avatars offer more identity and performance flexibility than traditional live streaming where one uses their own body, little research has focused on how this flexibility influences how creators present themselves. This research thus seeks to fill this gap by presenting results from a qualitative study of 16 Chinese-speaking VTubers' streaming practices. The data revealed that the virtual avatars that were used while live streaming afforded creators opportunities to present themselves using inflated presentations and resulted in inclusive interactions with viewers. The results also unveiled the inflated, and often sexualized, gender expressions of VTubers while they were situated in misogynistic environments. The socio-technical facets of VTubing were found to potentially reduce sexual harassment and sexism, whilst also raising self-objectification concerns.
\end{abstract}



\received{January 2023}
\received[revised]{July 2023}
\received[accepted]{November 2023}

\maketitle


\section{Introduction}
Live streaming is a new form of interactive media that combines the real-time broadcasting of high-fidelity audio/video stream with low-fidelity text-based IRC chat channels~\cite{Hamilton2014}. Prior research has shed light on how live streams function as a virtual third place \cite{Oldenburg1999} that has reshaped online social interactions among streamers and viewers ~\cite{Hamilton2014,Lu2018streamwiki,Lu2018youwatch,Cai2019,Hamilton2018,Wohn2018} due to its unique technological affordances. Furthermore, live streaming has also been found to be largely male-dominated~\cite{Wotanis2014,Doring2019,Freeman2020}, where female streamers were grossly under-represented and constantly harassed.

The adoption of a virtual avatar, however, can enable live streamers to create a new virtual identity that provides relative anonymity, affords alternate streaming performance experiences, and can enable for increased self-presentation opportunities including gender expression. The practise of using a 2D or 3D virtual avatar that is animated and voiced by a human actor (i.e.,  `Nakanohito') is called Virtual YouTubing or VTubing. Since 2016, live streaming and video sharing platforms such as YouTube, Twitch, and Bilibili (a Chinese video-sharing and streaming platform) have seen an rapid increase in the number of Virtual YouTubers or VTubers that create and contribute content. Typical VTubers usually maintain a virtual identity to engage with viewers while live streaming, and often share singing or dancing videos using their virtual identities.
Unlike traditional streamers, VTubers, notably in Asia, were known to be predominantly female. Prior work \cite{Lu2021} from the viewer side also unveiled the prevalence of sexism and misogyny among audiences that migrated from the Otaku subculture \cite{ito2012fandom} and virtual idol industry. How VTubers, notably female VTubers, present their identities via a virtual avatar, remains a significant yet unanswered question.

Before the emergence of VTubers, several researchers explored the opportunities that virtual avatars provided for self-presentation in avatar-based systems such as online games (e.g., \textit{World of Warcraft}) and social Virtual Reality (VR; e.g., \textit{VRChat}). This line of research highlighted how the \textit{proteus effect} \cite{Yee2007Proteus,Yee2009}, i.e., a phenomenon where user behaviour could be influenced by the characteristics of an avatar, was a key aspect of self-presentation when using virtual avatars.
While such systems have enabled users to experiment with new identities that are difficult to display in the physical world ~\cite{Stets2000,Huh2010,Bessiere2007,Kim2012},
user agency is controlled and limited by the choices made by the designers of avatar-mediated systems~\cite{Kolko1999,Mcdonough1999}, including affordances of Character Creation Interfaces (CCIs) that support limited flexibility~\cite{Ducheneaut2009, Mcarthur2015} and rules of the virtual world that often dictates platform-specific self-presentation \cite{Freeman2015,Freeman2016,Freeman2021}. VTubers can overcome the constraints of CCIs or virtual worlds by using design software to endow their avatars with complex human characteristic (e.g., rich facial expressions), and fine-tune their identities through live streaming for a long time beyond a one-shot customization. They can also actively engage and interact with viewers using their customized identities during or beyond live streams (e.g., Discord) without the constraints of a virtual world.

Among previous literature, it is still unclear how customizable virtual avatars in live streams afford self-presentation, influence streaming practices, and how streamer agency is configured while VTubing. Thus, we conducted semi-structured interviews with 9 VTubers and gathered observations of archived videos and live streams from 7 VTuber channels to answer the following research questions:
\begin{itemize}
    \item \textbf{RQ1}: How do VTubers construct virtual identities and what differences are there to real-person streamers?
    \item \textbf{RQ2}: How are streamer performances reshaped when a virtual identities are adopted?
    \item \textbf{RQ3}: How do VTubers perform their virtual gender identities (predominantly female) during live streams?
\end{itemize}

The study data revealed the strategies that streamers use to customize and manage virtual identities that contain elements of their real self, and how such identities afford more exaggerated performances and more candid and inclusive social interactions between viewers and steamers, among others.
The contributions of this research to HCI and CSCW are thus
(i) a nuanced understanding of how VTubers customize their virtual identity and engage with their audiences, as well as how viewer-streamer interactions are thus reshaped or extended, (ii) an exploration of the unique performance mechanisms that are used when virtual avatars are deployed while live streaming, which enhances our understanding of avatar-mediated self-presentation in social computing systems and
(iii) the unique gender expressions of predominantly female VTubers when they are exposed to misogynist environments, which opens up opportunities for future CSCW research.
\section{related work}
The present research was motivated by prior work on self-presentation in online social spaces and  avatar-based systems. In addition to reviewing literature from this area, we also review recent work on live streaming as a platform for self-presentation, focusing on gender and sexual identity.

\subsection{Self-Presentation in Social Media}
Self-presentation is a long-standing topic in communication research that dates back to the 1950s when Erving Goffman conceptualised everyday life as a theatrical performance \cite{Goffman1978}. Goffman argued that there is a `front stage' in social interaction where, similar to theatrical performances, individuals attempt to manage the impression each audience member forms about them. Compared to traditional face-to-face communication, impression management online is further complicated by socio-technical systems due to audiences of different norms and expectations being collapsed into the same context \cite{Marwick2011}. Hogan argued that in a collapsed online context, users tend to reduce their presentation to the `lowest common denominator' \cite{Hogan2010}. He also introduced an `exhibition' approach to analyze online self-presentation, which distinguishes `performances' from the `exhibition'. The performance stage takes place in real time, while an asynchronous `exhibition' stage enables users to manage and redistribute the `artifacts' that are created during the performance stage.

Drawing upon Goffman's theatrical metaphor and Hogan's exhibition approach, the previous literature has studied the technological affordances of online social media platforms~\cite{Zhao2013,Devito2017} and how users selectively crafted their identities~\cite{Kairam2012,Xiao2020} based on `the imagined audience'~\cite{Litt2012,Litt2016}. For example, Zhao et al. painted a multi-faceted picture of Facebook usage, arguing that users experienced the platform as three functional regions, i.e., a performance region for managing recent data, an exhibition region for long-term self-presentation, and a personal region that acted as a personal archive or diary \cite{Zhao2013}. In a study of secondary accounts (finsta) on Instagram, Xiao et al. explained the usage of such accounts as `intimate reconfiguration', where predominantly female users repurposed an existing socio-technical platform to present a non-serious, messy image of themselves to friends  \cite{Xiao2020}.

Similar to many other streamers, VTubers build communities using social media such as Discord beyond streaming platforms. Our study provided insights into how VTubers construct, maintain, and perform a pseudonymous virtual identity across live streaming and social media platforms.

\subsection{Self-Presentation in Avatar-Mediated Systems}
Apart from those intensively-studied social media platforms, another body of research has focused on avatar-based systems such as massively multi-player games, virtual worlds, social VR, and so on. As Manninen et al. posited, the use of a virtual avatar as a proxy for interaction in the virtual environment is a major difference between virtual worlds and the physical world ~\cite{Manninen2007}. Prior research has thus studied user motivations and practices for managing avatar-mediated identities in various virtual worlds \cite{Birk2016,Lin2014,Neustaedter2009}. For example, Neustaedter et al. found that Second Life users needed to balance pressure from in-game societal norms with their need to create a desired avatar \cite{Neustaedter2009}. Neustaedter et al. also identified four types of user identity needsthat presented unique challenges for avatar design (i.e., Realistics, Ideals, Fantasies, and Roleplayers). Other research has focused on the formation of group identity and its implications in virtual spaces \cite{Yu2008,Park2013}.

The presentation of one's self via a virtual avatar in a virtual world exemplifies the `proteus effect' \cite{Yee2007Proteus}, where users are prone to altering their self-presentation based on inferences of their avatars' expected behaviours. Yee et al.'s research found that such behavioural changes also translated to real-life face-to-face communication \cite{Yee2009}. Later research identified that users form attachments~\cite{Van2012,Bopp2019,Birk2016} to their virtual avatars, especially when an avatar had a sense of personality, unique behaviours, intentions, and style. This effect suggests that a user might see their avatar as a second self in the virtual world that they would like to protect and may worry about~\cite{Schroeder2012}, or become concerned if it was under attack or died~\cite{Wolfendale2007}. More importantly, virtual avatars have been found to afford different social values and norms than those held in in real life. For example, Freeman et al.'s studies on in-game marriage in \textit{Audition} demonstrated how the gender of an avatar in the game impacted users' self-presentation during intimate experiences \cite{Freeman2016,Freeman2015}. They found that avatar-mediated relationships featured explicit performances of users' self-development, sexuality, and ethnicity in the physical world. Based on data collected in \textit{Second Life}, Yee et al. found that virtual avatars transferred gender impact on interpersonal distances in the offline world into virtual environments and that intimate virtual interactions still supported Equilibrium Theory \cite{Argyle1965} in real life \cite{Yee2007Intimacy}. In addition, virtual avatars enabled users to experiment with completely new identities (e.g., cross-gender play~\cite{Freeman2015,Freeman2016,Huh2010}), or acted as an reaffirmation of users' existing identities~\cite{Yee2011,Ruberg2017,Haimson2018}. Ducheneaut et al. conducted an online survey of more than a hundred participants in 2009 and found that users of \textit{Maple Story}, \textit{World of Warcraft}, and \textit{Second Life} were motivated to create digital bodies that were different from their own to realize an idealised self, stand out, or follow a trend~\cite{Ducheneaut2009}.

In these avatar-based systems, however, it is the designers that ultimately restrict
 the agency of a user~\cite{Kolko1999,Mcdonough1999}. Although a user's virtual avatar can be customized via a Character Creation Interface (CCI), these interfaces support limited flexibility~\cite{Ducheneaut2009, Mcarthur2015}. Additionally, the performances of an avatar are also limited and affected by the design of the virtual world. For example, a study of social VR found self-presentation of users to be platform-specific~\cite{Freeman2021}, based on the social atmospheres afforded by the platform.  Freeman et al.'s  research on in-game marriage also observed user adaptations to the heterosexual view of marriage in \textit{Audition} ~\cite{Freeman2015,Freeman2016}. How avatar-mediated presentation would change if users, rather than platform designers, dictated identity customisation and performances remains an open research question. The present research aims to fill this gap by investigating the configuration of virtual avatars in live streaming settings, where creators have much more freedom.

\subsection{Live Streaming as a New Medium of Self-Presentation}
Since the late 2000s, live streaming has become increasingly popular across social media platforms and has been leveraged by users to build online communities and form social relationships ~\cite{Taylor2018,Consalvo2017}. The affordances of live streaming have been intensively researched from the perspective of performance and interaction~\cite{Pellicone2017,Kempe2019,Cai2019,Wohn2018}, cultural engagement~\cite{lu2019feel,lu2021streamsketch}, content moderation~\cite{Cai2019,Seering2017}, privacy management~\cite{Li2018,wu2022concerned}, and so on. Prior work has also found that the interactivity and sociality of live streaming has contributed greatly to its popularity~\cite{Hamilton2014,Lu2018youwatch,Wulf2020,lu2019vicariously}. 

In such contexts, authenticity has been highlighted as one of the most important features influencing how live streamers present themselves. For example, Pellicone et al. found that the self-presentation of gaming live streamers was related to the need to build a stronger community of viewers and that a key skill to doing so was to develop an authentic and natural persona that reflected one's true self, while also communicating a unique attitude toward gameplay \cite{Pellicone2017}. In a study of E-commerce live streaming, Tang et al. found that rural female streamers in China regarded authenticity as key to engaging viewers and that their self-presentation was centred around the `rurality' and `rawness' of their lives \cite{Tang2022}.
Wu et al. further identified the tensions between preserving privacy and constructing attractiveness through disclosure of one's authentic self \cite{wu2022concerned}. Wu et al.'s study provided a nuanced description of how streamers leveraged both platform-supported and self-developed strategies to manage the disclosure of private information.

It is also worth noting that live streaming platforms are known to be male dominated and are often criticized for sexism and harassment against female streamers \cite{Wotanis2014,Doring2019,Freeman2020}. Focusing on the most-subscribed female YouTubers in 2013, Wotanis et al. found that female creators were under-represented and more vulnerable to negative feedback (i.e., hostile and sexist) than their male counterparts ~\cite{Wotanis2014}. This was further replicated in a follow-up study of the 100 most subscribed YouTube channels in nine different countries and a content analysis of thousands of video comments~\cite{Doring2019}. To understand how female streamers managed their identities and presented themselves, Freeman et al. conducted semi-structured interviews with 25 female and LGBTQ streamers ~\cite{Freeman2020}. They found that the digital representations of streamers were self-driven rather than audience/performance-oriented, and that such representations acted as an affirmation of one's perception of self and as empowerment to advocate for equity in the live streaming community.

The present research contributes to this line of inquiry by unveiling the affordances of avatar-mediated live streaming, which brings about pseudonymity but preserves authenticity. This research also highlights the inflated and often sexualised gender expressions that are employed by live streamers and the generally inclusive environments of viewers. Gender-related findings and implications are also examined through lens of feminism.

\section{Methodology}
To gain deeper insights into VTubers' identities, streaming practices, and self-presentation from third-person perspectives and creators' subjective and conscious reflections, we conducted participant observation of 7 VTuber channels and semi-structured interviews with 9 VTubers (C1-7, I1-9; \autoref{VTuber_Demographics}).
The 7 observed VTuber (C1-7) and 9 interviewed VTubers (I1-9) are independent of each other.
As VTubing typically involves both video sharing and live streaming, all VTubers involved also regularly uploaded videos to their channels of themselves singing or dancing, highlights of their streams, etc. All VTubers were ethnically Chinese and regularly live streamed via YouTube Live, Twitch, or Bilibili.
Bilibili is a Chinese video sharing platform similar to YouTube, that provides live streaming features and is famous for its Japanese ACG (anime, comics, and gaming) culture where VTubing originated.

\begin{table*}[htb]
    \centering
    \renewcommand{\arraystretch}{1.2}
    \resizebox{\textwidth}{!}{\begin{tabular}{ccccccccc}
        \toprule
         \textbf{VTuber ID} & \textbf{Platform(s)} & \textbf{Language} & \textbf{Region} & \makecell[c]{\textbf{Avatar} \\ \textbf{Gender}} & \makecell[c]{\textbf{VTuber} \\ \textbf{Gender}} & \makecell[c]{\textbf{Affiliated with} \\ \textbf{a Company}} & \makecell[c]{\textbf{Number of} \\ \textbf{Subscribers}} & \textbf{Evaluation Method}\\
         \midrule
         C1 & YouTube & Cantonese & Hong Kong & Female & - & No & 10K-50K & Observed\\
         C2 & Bilibili & Mandarin & China (mainland) & Female & - & Yes & >500K & Observed\\
         C3 & YouTube & Mandarin & Taiwan & Female & - & Yes & 50K-100K & Observed\\
         C4 & Bilibili & Mandarin & China (mainland) & Female & - & Yes & >500K & Observed\\
         C5 & Bilibili & Mandarin & China (mainland) & Male & - & Yes & 50K-100K & Observed\\
         C6 & YouTube & Mandarin & Taiwan & Male (Animal) & - & Yes & 10K-50K & Observed\\
         C7 & YouTube & Cantonese & Hong Kong & Male & - & No & 2.5K-10K & Observed\\
         I1 & YouTube & Cantonese & Hong Kong & Female & Female & No & 2.5K-10K & Interviewed\\
         I2 & Bilibili & Mandarin & China (mainland) & Female & Female & No & 50K-100K & Interviewed\\
         I3 & YouTube & Cantonese & Hong Kong & Male & Male & No & 2.5K-10K & Interviewed\\
         I4 & YouTube \& Twitch & Mandarin & Taiwan & Female & Female & No & 2.5K-10K & Interviewed\\
         I5 & YouTube & Mandarin & Taiwan & Female & Female & No & 10K-50K & Interviewed\\
         I6 & Bilibili & Mandarin & China (mainland) & Female & Female & Yes & 10K-50K & Interviewed\\
         I7 & Bilibili & Mandarin & China (mainland) & Female & Female & Yes & 2.5K-10K & Interviewed\\
         I8 & Bilibili & Mandarin & China (mainland) & Male & Male & No & 10K-50K & Interviewed\\
         I9 & Bilibili & Mandarin & China (mainland) & Female & Female & Yes & 50K-100K & Interviewed\\
         \bottomrule
    \end{tabular}}
    \caption{Demographics of the VTubers observed and interviewed in our study. Note that in the Participant Gender column, '-' denotes that the gender was not available.}
    \label{VTuber_Demographics}
\end{table*}

\subsection{Interviews}
Purposive sampling \cite{Robinson2014} was adopted during the recruitment of interviewees to ensure that participants came from diverse cultures, genders, and platforms. We reached out to approximately one hundred VTubers through e-mail or direct messaging, according to recommendations of streaming platforms and fan groups. All VTubers approached streamed regularly every week and had streamed for at least a month at the time they were contacted.
A total of nine VTubers (6 female) agreed to be interviewed.
The interviewees had live streamed as a VTuber for at least half a year and accumulated at least 3K subscribers at the time they were interviewed.

Interviews were conducted remotely in a semi-structured manner via online meeting software (Zoom/Tencent) or audio calls (WeChat/Discord) in November and December of 2021, except the interview of I9, which was done via QQ chat (a Chinese instant messaging app). The interview session lasted between 40 to 60 minutes and the interview questions were focused on the following themes: interviewees' motivations for becoming a VTuber (e.g., ``How did you become a VTuber?''), the construction of their virtual identity (e.g., ``How did your virtual identity evolve through live streaming?''), their live streaming practices via an avatar (``Describe how you performed your identity during live streaming'', ``What software did you use?'', etc.), and so forth. For each theme we initially prepared three or more questions as a starting point of our interview. Additionally, before each interview, the researchers browsed recent videos of the VTuber and reviewed their social media posts (e.g., Twitter, Instagram, Discord, etc.) to familiarise themselves with the interviewee and create interviewee-specific questions about their identity or performance (e.g., ``Why did you set your identity to be an animal?'', ``How did you come up with the background story of your identity?'', etc.). All interviewees were also asked to recall their most memorable viewer feedback, the greatest challenge of being a VTuber, and the biggest difference between being VTubers or real-person streamers.

In particular, interviews with the two Cantonese-speaking VTubers (i.e., I1, I3) were conducted in Cantonese by a native-speaking researcher, while the others were conducted in Mandarin. The two Cantonese-speaking VTubers joined the Zoom meeting as their animated virtual avatar, while I1 performed as her virtual character (e.g, by raising their voice pitch, changing their avatar's clothes, etc.) while being interviewed.
Though performing virtual identities, all interviewees agreed beforehand to answer  questions on behalf of their real selves (the `Nakanohito') instead of virtual identities.
All interview sessions were recorded with consent of the interviewee and later transcribed into Mandarin.

\subsection{Observations}
Candidate VTubers channels were selected from public ranking websites, such as Playboard \footnote{https://playboard.co/en/}, or via the recommendations of platforms (e.g., the VTuber section of Bilibili). The seven VTubers that were selected had between 2.95K to 818K subscribers. Five were affiliated with a company (the other two claimed to be individual VTubers) and four were female. Their data compensated for the bias in the interviewee sampling, since VTubers with smaller viewership were more inclined to be interviewed.

The observation of the seven channels were assigned to four researchers (all with a background in social media research), with each being responsible for one or two channels. Prior to the observation process, the four researchers engaged in a discussion to develop a coding scheme. They agreed to focus on the following aspects of stream data:
\begin{itemize}
  \item How an avatar was designed, placed, rigged, animated, and voiced by the VTuber, and what he or she said behind the avatar
  \item What kind of an identity was implied via a VTuber's performance or what he or she said about his or her identity
  \item Viewer feedback and comments so that we could understand how viewers reacted to the VTuber and how the reactions affected performances
\end{itemize}
Aside from emerging themes related to the above three aspects, the researchers also agreed to provide descriptions of the data (e.g., what animation was played, how was the avatar voiced, what viewers reacted exactly, etc.), include links to saved streams or videos along with timestamps, capture viewer comments, and include screenshots if necessary to reconstruct the scene at that particular moment.

Following the above coding scheme, we first watched and coded the `debut stream' of each VTuber, which was usually the first stream of the VTuber's channel where the VTuber introduced themselves. We then joined their community via their QQ group or Discord, and followed their streaming schedule to watch and code each subsequent stream. The observation sessions took place in November 2021 and lasted for at least a week for each VTuber. During this period, we watched all their live streams and documented initial emerging themes. In total at least 5 live streams were coded for each VTuber, each stream no shorter than 2 hours. For each VTuber, we also coded their five most watched videos as ranked by the platform at the end of each observation session. For C2, the first author saved C2's most recent social media (Weibo) posts for the week, because she and her viewers often mentioned her social media posts during streams. The four researchers then held a discussion to cross-check codes and themes that had emerged to ensure their  reliability and consistency. To achieve consensus, each researcher shared and explained their codes, using the saved videos as references, while the others posed questions and provided feedback. The first author took notes of each confirmed code or theme after a collective agreement was reached.

\subsection{Data Analysis}
The data obtained from the qualitative study was comprised of interview transcripts and initial codes from the observed videos and live streams. We then utilized an open coding method~\cite{Corbin2014} to analyze the data in two phases. In the first phase, the four researchers individually coded all interview transcripts and observed video and live stream codes into emergent themes and categories, such as identity construction, performance mechanisms, viewer-streamer interaction, gender expression, etc. The initial codes of observed videos and live streams consisted of initial themes with timestamps, descriptions, and viewer responses and screenshots. By analyzing this data collectively, the events could be reconstructed visually at each timestamp. For example, at \underline{0:04:19} (timestamp) of stream \#2, C1 was \underline{overacting by pretending weeping} (initial theme) \underline{using a soft voice and a weeping animation} (description), but \underline{viewers taunted her} (viewer response), and a \underline{screenshot} was captured at that moment containing both video stream and text chat. All emergent themes (in this data analysis phase) in observation data were extracted based on the recreated events.

In the second phase, the resulting codes and themes were re-examined individually by each coder in an iterative manner: observation notes and interview transcripts were compared simultaneously, where themes emerging from transcripts were re-examined and merged with those in the observation data, and vice versa. During these two phases, each researcher maintained an individual codebook. The four researchers later held a review session to exchange and discuss all their codes to reach a consensus. All codes in each codebook were discussed one by one, and then merged with each other once an agreement was reached. The first author then merged all the confirmed emerging themes and codes in each researchers' codebook into a single document. Subsequently, during a conversation involving the first author (male) and the other two female researchers, all gender-related codes and themes were confirmed without any researcher raising doubts. The codes, themes, and related data were then translated into English for reporting.
\section{Findings}
Based on the interview and observed video and live stream data themes that emerged, we first report how VTubers constructed their virtual identities and how they performed using these identities during live streams. We then report on the unique gender expression of VTubers.

\subsection{The Construction of a Virtual Identity}
\subsubsection{Construction Process}
Before making their debut, VTubers first need to design their virtual avatars. Today, this is accomplished using design software such as VRoid Studio or Live2D. I1 (female avatar \& Nakanohito), for example, said she hired artists to design her an avatar based on features such as `14 year old', `cute', `talkative', etc. Some VTubers with designer or drawing expertise reported that they designed or learned to design their avatars themselves. For example, I5 (female avatar \& Nakanohito) said she used to be a game designer and had an art major, so she designed most of her avatar elements and streamed content. I3 (male avatar \& Nakanohito) said his avatar was initially designed by himself, but that he later hired artists to refine his avatar based on elements he had designed on his own. 

When a VTuber channel debuts, there will usually be an initial stream during which the VTuber introduces the identity of their avatar and demonstrates its upbringing, personality, hobbies, and so on, which partly indicates how the avatar will be performed by the `Nakanohito'. After the `debut' stream, the virtual identity is further enriched and fine-tuned, both in its visual design and in how it is performed by the Nakanohito. For example, we observed that C1 and I5's avatars put on a new Christmas-style outfit while streaming on Christmas Eve. I1 (female avatar \& Nakanohito) reported that she once designed summer clothing by herself using Live2D because she ``\textit{would like to change clothes with her fans in reality when the weather turns hot}'' (\autoref{sparrow_summer}).  She also claimed to have deliberately raised the pitch of her voice after streaming for a while because she felt it ``\textit{sounds cuter}'' and ``\textit{matches the 14-year-old identity}''. I2 (female avatar \& Nakanohito) said that her virtual identity, such as they way she dressed, her accessory choices, and her avatar's personality was partly determined after discussions with her fans during live streams.

Besides livestreaming, today's VTubers also upload videos to YouTube or Bilibili to archive their identity development and performance highlights via debut streams, self-introduction videos, outfit or costume reveal streams, stream highlights, singing, dancing, etc. Some VTubers also ran social media accounts using their virtual identities. For instance, C1 and I1 (both female avatar) used Instagram to post stream highlights and identity developments (e.g., outfit reveals, new 3D models, etc). C2 (female avatar) also used Weibo (a Chinese microblogging website) as an archive and to share her personal life.

\subsubsection{Blended Virtual and Real Identities}
VTubers usually performed in a way that matched the expected visual and behavioral characteristics of their avatars, which differs from real-person streaming, which emphasises authenticity \cite{Pellicone2017,Tang2022}.
For example, C4's (female avatar) virtual identity was to be an AI so she often acted as if she were an AI that spoke like a robot, switched between languages, and ridiculed the limits of humans. However, even when using such extreme personas, we found that such identities tended to blend with the real self of the VTuber. The design of the virtual avatar, along with its made-up virtual world and background, usually borrowed ideas from the real life or idealisations of the VTuber. For example, I1's (female avatar \& Nakanohito) avatar was designed to resemble a sparrow because it was quite commonplace in Hong Kong and she wanted to promote the culture of Hong Kong (\autoref{sparrow}). According to I2 (female avatar \& Nakanohito), her identity used to be a herb fairy (\autoref{dolores}) in a virtual world. She said she was not in good health at that time, so she claimed her identity was capable of curing all illnesses.

Apart from their design, of all our interviewees, I2, I4, I6, and I9 (
4 out of 9
) emphasised how they performed their avatar to mimic themselves in real life., e.g., \textit{``Actually I did not deliberately perform a completely made-up identity. On the contrary, I perform the identity with my personalities in reality. I hope my viewers like my real self rather than just the avatar alone.''} (I4, female).
I4, I8, and I9 (
3 out of 9
) also mentioned that it could be difficult to maintain a virtual identity that differed greatly from a VTuber's own identity in the long run. I4 (female avatar \& Nakanohito) recalled that she acted more naturally and closer to herself after the first week of streaming, i.e., \textit{``At the very beginning, I thought I might need to perform the avatar in a (more feminine) way that matches its charismatic look but after a week, I felt being not true to yourself is exhaustive.''}
I1 and I6 (both female avatar \& Nakanohito), however, reported using a higher pitched voice than usual when voicing their avatars but claimed that the personalities of their virtual identities were generally similar to themselves.

Furthermore, for VTubers, the construction of their virtual identity was usually not the end goal, but a means to present part of their real selves. VTubers were motivated to use their avatar to stream what they liked and convey their genuine thoughts instead of acting only as the virtual identity. For example, during her interview, I6 (female avatar \& Nakanohito) stressed that \textit{``When streaming as a VTuber, I would not maintain a very complicated identity. Perhaps to my mind, it is [just] used for the convenience of conveying what I want to [stream]... I hope my audience focuses on my streamed content [especially singing streams], more than the avatar itself.''} 
I8 also mentioned that he aimed to share his favourite history knowledge while being a VTuber. He said he would strive to improve his history streams, although some fans might find them unappealing.

\subsubsection{The Pseudonymity Afforded by Virtual Identities}
Although virtual identities were found to mirror the VTuber in real life, most interviewees stressed the importance of the pseudonymity that such identities offered, which they believed was a major difference between themselves and real-person streamers. Such pseudonymity implies a pseudonymous identity as a brand of their streams and a relative anonymity where no personally identifiable information would be mentioned or talked about. I2 (female avatar \& Nakanohito) said that, \textit{``I make efforts to maintain my virtual identity. To many of my fans, I'm a considerate friend but it would cause a sense of incongruity among viewers if they knew too much about my privacy in real life.''}
Besides the `sense of incongruity', this steamer also noted that some streamers would specifically choose to become VTubers for privacy and anonymity. I8 also mentioned that VTubers generally do not like to be identified to avoid complications in real life.

Furthermore, interviewees found that the benefits of such pseudonymity have empowered many to become streamers, especially women. For example, I2 (female avatar \& Nakanohito) used to be a real-person streamer before being a VTuber. She said she started streaming at the age of 15 or 16 and claimed to be sexually harassed quite regularly. After becoming a VTuber, she said, \textit{``I've rarely been harassed since then, though sporadic cases exist. Before that I felt the whole environment was hostile against women.''}
This sentiment was also shared by I7 (female avatar \& Nakanohito), who explained in retrospect that, \textit{``They [fans of real-person streamers] kind of have a sense of possession .. they feel like paying to adopt the streamer [as if she was their girlfriends]. I know of a friend, she happened to post a selfie with her boyfriend and her fans really exploded. You know what I mean? ... however for VTubers, such cases are very rare.''}

Apart from the hostility towards women, I1 (female avatar \& Nakanohito), I3 (male avatar \& Nakanohito) and I4 (female avatar \& Nakanohito) all noted that being a VTuber``\textit{it would save the hassle of applying makeup}'', which potentially lowers the barrier of becoming a streamer. I7 and I9 (both female avatar \& Nakanohito) also mentioned that for real-person streamers ``\textit{appearance is one of the most important factors that attract viewership}'' and a (female) streamer probably has to be good-looking. However for VTubers, ``\textit{what appeals to viewers could be your avatar design, your voice, or your personality, etc., instead}''. I2 (female avatar \& Nakanohito) also added that a virtual avatar could ``\textit{act as a channel of expression for streamers too shy to perform publicly}''. I3, for example, was a male VTuber who regularly streamed female-oriented ASMR. He emphasised that he actually ``\textit{could not do the stream without the virtual avatar identity, because it would be very embarrassing if it would be real-person streams}''.

\hlwq{
\begin{figure*}[ht]
\begin{subfigure}{.8\textwidth}
    \centering\captionsetup{width=\linewidth}
    \includegraphics[width=\linewidth]{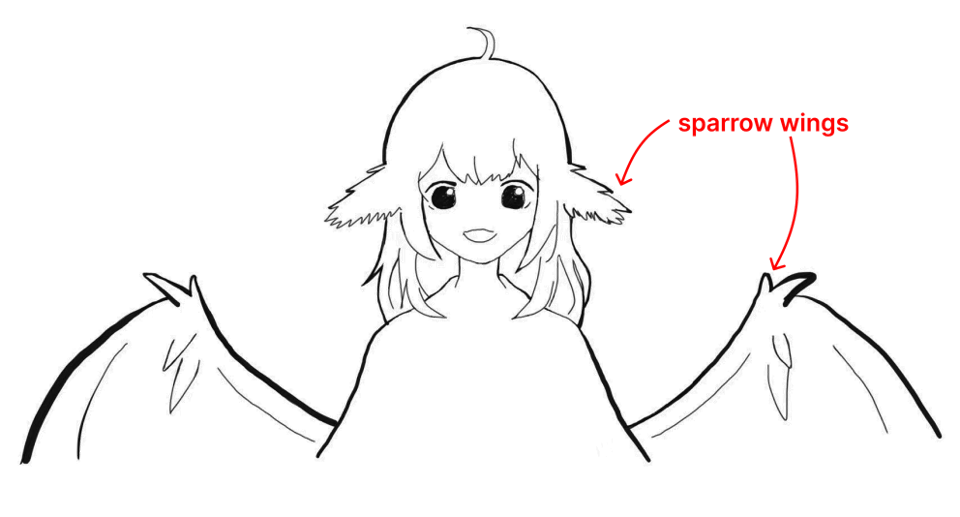}
    \subcaption{The identity of I1 was initially a 14-year-old and sparrow-looking girl. I1 hired an artist to design the avatar based on keywords such as 14-year-old, cute, talkative, etc. The avatar used accessories resembling sparrow wings to represent Hong Kong.}
    \label{sparrow}
\end{subfigure}
\begin{subfigure}{.45\textwidth}
    \centering\captionsetup{width=.85\linewidth}
    \includegraphics[width=\linewidth]{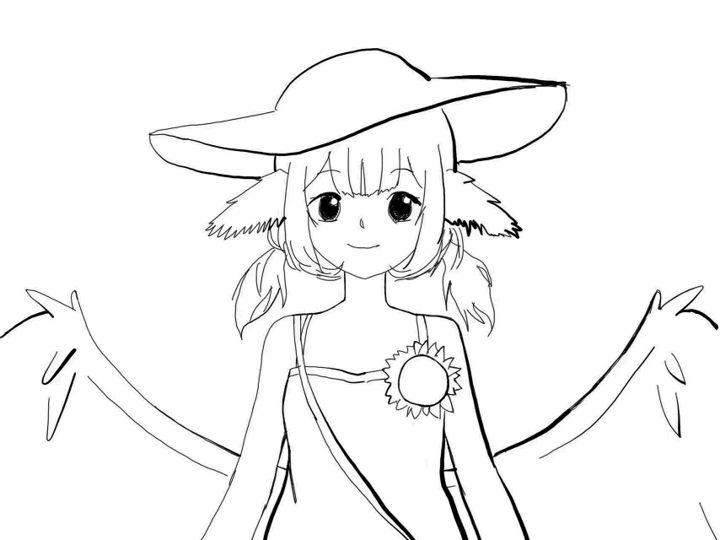}
    \subcaption{I1 wanted to wear summer clothes with her fans in reality so she designed a summer outfit using Live2D.}
    \label{sparrow_summer}
\end{subfigure}
\begin{subfigure}{.5\textwidth}
    \centering\captionsetup{width=.9\linewidth}
    \includegraphics[width=\linewidth]{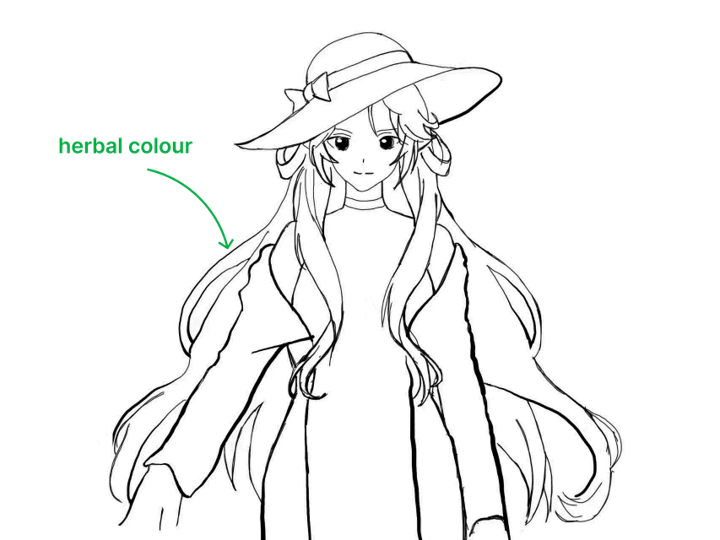}
    \subcaption{The identity of I2 was said to be a herb fairy that was capable of curing all illnesses. The avatar design, accessories, and personality was determined via discussion with her fans.}
    \label{dolores}
\end{subfigure}
\caption{Visual representations of participant VTubers' identity constructions.
The three sketches were hand drawn by the research team based on observed avatar appearances.
}
\end{figure*}
}

\subsubsection{Summary}
These findings revealed that the construction of a VTuber identity requires design effort before one's debut and long-term fine-tuning inspired by viewer feedback to increase viewer engagement. Such identities were virtual but blended with one's real self and afforded pseudonymity. The major differences between VTubers and real-person streamers were two-fold. First, although authenticity was required for both, the identity of a VTuber was virtual, i.e., VTubers initially performed their virtual identity to present more of their real and authentic self. Secondly, although VTubers needed to build an identity as a brand, similar to a real person, such an identity did not involve any personally identifiable information and offered relative anonymity.

\subsection{Avatar-Mediated Performances}
The observations and interviews also honed in on several facets of avatar-mediated performances such as the streaming practices of VTubers, how VTubers act during their streams, the limitations existing technologies impose on VTubers, and the inclusiveness of viewer-streamer interaction.

\subsubsection{Streaming Practices When Performing as an Avatar}
Due to the absence of a real person, VTubers often need to `rig' their avatar. In the live streams and videos we observed, this rigging included changing the clothing, accessories, and facial expressions of the avatar, and triggering animations of the avatar in real-time via a keyboard or motion tracking using software such as VTube Studio. In addition to the avatar, the streamed background content was comprised of 2D resources such as sketches, drawings, images, or shared screens of games. An exception was C7 (male avatar), who streamed within a 3D virtual world using a 3D avatar that was animated via full-body motion tracking. It is also worth noting that 2D resources are often used as illustrations while the VTuber is performing (e.g., telling a story or making a point), instead of as the main streaming content. For example, in the debut stream of C6 (male animal avatar), he told background stories about his virtual avatar while drawing illustrations on a shared screen in real time. C2 (female avatar) was also observed introducing her avatar while presenting 2D slides with sketches that were paired with text illustrations.

    

For VTuber viewers the inclination to participate and interact with VTubers' virtual identity contributed to their motivation to watch a stream. Therefore, like many other streamers \cite{Taylor2018,Hamilton2014,Faas2018}, VTubers build communities by hosting Discord servers or QQ groups and try to involve their fans in the stream as much as possible. In comparison with real-person streaming experiences, I2 (female avatar \& Nakanohito) emphasised that her fans became more attached to her, i.e., \textit{``I used to stream League of Legends and my viewers at that time like the game itself more than the streamer they watch ... so they would only watch the stream if you are a good player, and leave the stream if you switch to another game ... now I feel my viewers are more attached to me (the virtual identity) than before, whatever I stream.''} She also recalled an instance when she fell asleep while streaming and her fans tried contacting her friends for her safety, starting Shiritori (a text game) in the chat to show that they were not leaving her.
I4 (female avatar \& Nakanohito) also agreed about such attachment during her interview and emphasised that many of her fans ``\textit{came to watch the stream because they would like to interact with the streamer}'' and that is why her viewership may have dropped when she switched from chatting to gaming because she ``\textit{could not focus on interacting with fans while playing games}''.

To engage with viewers, in the drawing streams of I5 (female avatar \& Nakanohito), whatever she drew was determined by the viewers in her chat. I3 (male avatar \& Nakanohito) also hosted polls in his Discord channel to let fans determine what he would perform in upcoming streams. During our observations, we noticed that chit-chat was the most common stream type.
During these streams, VTubers engage in casual and recreational conversations with all viewers in the chat about various topics.
I1 and I6 also reported that chit-chat streams were the most popular on their channels because they allowed for active fan participation.

\begin{table*}[htb]
    \centering
    \begin{tabular}{ll}
        \toprule
        \textbf{Screaming} & C1, C3, C4, C6, C7, I5, I7 \\
        \textbf{Swearing} & C1, C3, C6, I4, I7 \\
        \textbf{Telling porn jokes} & C1, C2, C3, I2, C6, I4, I5, I9\\
        \textbf{Singing Off-key} & C1, C3\\
        \textbf{Playing games poorly} & C1, C3, C5\\        
         \bottomrule
    \end{tabular}
    \caption{The ways VTubers intentionally exaggerated their performances during live streams.
    7 out of 7 observed VTubers and 5 out of 9 interviewees
    were found to exaagerate their performances}
    \label{VTuber_performance}
\end{table*}


\subsubsection{Overacting}
During VTuber performances, we observed various exaggerations in terms of voicing and acting. For example, in a self introduction, C3 (female avatar) claimed to have ``Multiple Personality Disorder'' so she sometimes voiced her avatar by speaking in a gentle and comforting manner, but at other times she would be vulgar or blast her viewers. I7 (female avatar \& Nakanohito) also stated that she would switch between sweet-sounding Japanese and heavily accented Chinese to exaggerate her avatar.

In addition to voicing, VTubers tended to exaggeratedly playact for dramatic effect. Intentionally screaming, swearing, telling porn jokes, singing off-key, or playing games poorly were common behaviors exhibited by most VTubers (\autoref{VTuber_performance}). As an example, C1 and C4 were often found to play cute, pretending to weep pitifully via a weeping animation and adopting an affected soft voice when they played poorly in a game or were taunted by viewers.

The reasons behind such overacting were in part due to the reduced cues available in avatar-mediated performances compared to real-person streamers (I4, I5 and I9). As an explanation of the higher-than-usual voice they used when voicing their avatars, I5 (female avatar \& Nakanohito) said that, \textit{``It would be straightforward for a real-person streamer to express their emotion, due to the presence of a webcam, however, a VTuber would seem somewhat dull if the emotion lies solely on what you see through the avatar itself so we must use our voices to inflate our performances.''} I4 (female avatar \& Nakanohito) also noted the reduction of non-verbal cues and stated that she would act more dramatically to make her performances `noticeable'. When asked to reflect on the exaggerations of performances, I9 (female avatar \& Nakanohito) explained that \textit{``Live-streaming via a virtual avatar is still limited in its form. Apart from the avatar (visual design) itself, how you act, and how you voice the avatar also act as a means to attract audiences so some VTubers will exaggerate their performances.''}

\subsubsection{Inclusive Interaction}
By speaking behind a virtual avatar, VTubers were generally more candid and straightforward. Some VTubers, e.g., C3 and C4, did not shy from addressing controversial and personal issues. During C3's chit-chat streams, she often shared her personal experiences or anecdotes to inspire her fans. During one stream, she was asked about a controversial incident she was involved in, and after asking her company for consent, she confessed her genuine opinions and admitted that she was at least partly to blame. Among our interviewees, I5 (female avatar \& Nakanohito) emphasised open-mindedness and liberalness as the most important features of her stream. She explained that, \textit{``(fans) like me because they can talk about anything. Some other streamers might have rules on what cannot be talked (politics, sex, etc), but I don't have any restrictions.''}

Despite VTubers being relatively outspoken and overacting, viewers have been found to be quite tolerant of such behaviors \cite{Lu2021}. I2 (female avatar \& Nakanohito) stressed the tolerance of her viewers compared to real-person streaming experiences in game streams. She said that, \textit{`` The audiences become more tolerant ... The viewers when I was a real-person streamer came to watch your stream because of the game instead of the streamer. They could blame you if you were bad, or simply made mistakes during the game, however, when streaming as a VTuber, my viewers would joke around inoffensively on the same occasion, and teach you how to play''}.
This also agrees with our observations. When C3 (female avatar) swore at her viewers during her stream, instead of taking offence, her fans joked around and asked her to swear at them (sometimes even harsher) for fun, knowing that C3 was acting.

\subsubsection{Limitations of 2D Space}
Based on our interviews, a major limitation of VTuber performances is the 2D space they perform within. I4 mentioned the inconvenience of presenting 3D content via 2D avatars. By comparison with real-person streamers, I4 (female avatar \& Nakanohito) said, \textit{``The virtual avatar limits your means of performance. Despite richness in facial expressions, all your movements below would not be detected. Real-person streamers can make use of their hand gestures and even rotate their webcam 360 degrees.''} She also mentioned that VTubers needed to remain seated upright for head motions to be detected. When asked about their most desired improvements regardless of technological and financial constraints, 5 of the 8 VTubers interviewed said they would like to have a 3D avatar. I1 (female avatar \& Nakanohito) said she would like to make vlog using a 3D avatar to introduce the street scenes of the city she was from, however, the expense of creating a 3D avatar, along with the technological cost of animating it, was too limiting.

\subsection{Gender Expression}
The avatar design and performances of `Nakanohitos' demonstrated how gender expressions became much more exaggerated and sexualized than with real-person streamers. Thus, we first introduce the gender demographics of the VTuber community and then explore phenomenon of exaggerated gender expression. We also discuss the underlying reasons or motivations for the gender expression as derived from the interview data.

\subsubsection{Gender Demographics}
Contrary to traditional streamers, which have been found to be male dominated~\cite{Doring2019,Wotanis2014,Freeman2020}, today's VTubers' virtual identities are predominantly female. According to statistics from third-party ranking sites such as Virtual YouTuber Wiki\footnote{https://virtualyoutuber.fandom.com/wiki/Virtual\_YouTuber\_Wiki}, by January 2022, nearly 80\% of today's VTubers were using female avatars (most Nakanohitos were believed to be female as well). This is usually because VTuber viewers mainly consist of ACG lovers, who are mostly male and favour feminine avatars voiced by a feminine (though not necessarily female) Nakanohito. Among our female interviewees, I4 and I5 disclosed that 80\% of their viewers were male and most of them were from the ACG community. I2 said that of the 156 fans in her QQ group chat, only 4 were female.

During their interviews, both I3 (male avatar \& Nakanohito) and I5 (female avatar \& Nakanohito) mentioned that being a male (avatar) VTuber was much more difficult than being a female (avatar) VTuber because viewers preferred feminine performances. To attract female fans, I3 streamed ASMR content that targeted female viewers (e.g., erotic talk for straight women). Therefore, viewers of I3 were mainly female. Another male interviewee, I8, who streamed gaming and history knowledge, said he had few female fans because they would not be interested in his content.

\begin{table*}[htb]
    \centering
    \begin{tabular}{ll}
        \toprule
        \textbf{Inflated avatar appearance} & C1, I2, I5 \\
        \textbf{Deep v-neck dress} & I1 \\
        \textbf{Leg harness or fishnet stockings} & C1, I2, I2, I5\\
        \textbf{Military uniforms} & I5 \\
        \textbf{Sexualised ASMR} & I3 \\
        \textbf{Sweet and soft voices} & C1, I2\\
        \textbf{Consoling fans} & C3, I2\\
        \textbf{Pranking and mansplaining} & C5, C6, C7 \\
        \bottomrule
    \end{tabular}
    \caption{The techniques VTubers used to inflate the gender expression of their avatar. In total
    9 out of 16 VTubers
    have inflated gender expression}
    \label{VTuber_gender}
\end{table*}

\subsubsection{Inflated Gender Performance}
Motivated by their avatar design, the performance of a VTuber's gender role was often exaggerated and sexualized (\autoref{VTuber_gender}). For a female avatar, the sizes of the bust/waist/hips tended to be inflated (e.g., I2, I5), the legs were thin and long, the eyes were round and big (C1), and the face shapes were optimised (e.g., optimally heart-shaped). The clothing design also included exaggerated and provocative elements such as a deep v-neck dress (I1) or a leg harness or fishnet stockings (C1, I1, I2, I5). The masculinity of male VTubers was also inflated by the use of military uniforms (I8). As an exception, C5 experimented with avatar designs that differed from traditional gender roles (e.g., a male-presenting avatar voiced by a male Nakanohito that had common feminine design traits such as long hair, eye shadow, and blush.)

The gender characteristics of an avatar were further strengthened by the performance of the Nakanohito. Female VTubers tended to voice their avatars using softer, higher voices than usual to sound appealing, while male VTubers used much lower voices. In the female-oriented ASMR stream of I3 (male avatar \& Nakanohito), the voicing was sexual and included kissing imitations, heavy breathing, murmured sweet words, and so on. Gender stereotypes were also found to be somewhat extreme in VTubers' acting. Many female VTubers (C1, I1) acted cute by using sweet and soft voices and saying sweet words, or were gentle and considerate (C3-5) when consoling fans in the chat when they were down. Male VTubers, however, often acted naughty (C5, C6-7), and pranked or mansplained. In one game stream of C5, for dramatic effect, C5 acted over-confident and preached his strategies to his co-streamers by drawing illustrations, as if he were the team leader. These strategies did not end up working later in the game.

When asked to reflect on the performance of gender, many interviewees (I2, I4, I5) claimed that their gender expression was self-driven. For example, I4 (female) used an avatar different from herself in real life. She said she was more of a `boyish' and cool-looking girl in real life, however, her avatar was exaggerated into a much more feminine look. She claimed that her performance of the avatar's gender stayed true to her real self. She recalled that, \textit{    ``There might be a few new audiences that said, `You are a female VTuber. You should behave properly and not swear.' I would address them directly, `This is my style. You may please leave the channel if you don’t like me .' In this way, those that stay subscribed to you - most of them do - ended up those who accept you. My audiences actually like me this way since it is more authentic.''}
I3 (male), however, said his ASMR streams and sexual voicing were at least partly market-driven because he needs to build viewership.

\subsubsection{Sexual Suggestiveness}
We also observed the prevalence of sexual suggestiveness in VTubers' interaction with their viewers (C1-3, C6-7, I2-5,
9 out of 16 VTubers
). This included flirting, dirty talk, telling porn jokes, openly talking about sex or sex fetishes, etc. For example, in one chat stream of I2 (female), the VTuber said \textit{``Like you animals in the chat, however you deny being my puppies, how come you sound so aroused when addressing me as Master?''}
In another stream of C3 (female avatar), she once responded to a viewer that complimented her singing, saying \textit{``Your compliments make me almost come ... open your mouth, chat, I am about to come''}.

When asked to reflect on such sexual suggestions, I2 (along with I4 and I5) said her performances, even though sexually suggestive, were generally natural and spontaneous. I9 also mentioned flirtations in her interview and said she was neutral about it, because VTubers have to make up for the reduced cues of virtual avatars via their performances, of which flirting could be more effective. I7 (female avatar \& Nakanohito) compared such behavior to sexual suggestions in real-person streams. She claimed that real-person streamers might often ``\textit{turn to sexual appeal of their physical body to attract viewers}'', while ``\textit{it is impossible, at least visually, for a VTuber}''. She said that \textit{``a female (real-person) streamer once received a gift of 520 yuan (520 reads like I love you in Mandarin). She then made a hand heart directly upon her breasts in a very seductive way. For me (a VTuber), all I would do is to say thank you.''}
In her opinion, the performances of real-person streamers were much more sexualized and often took advantage of self-sexualization to attract viewers.
\section{Discussion}
These findings revealed a nuanced description of how VTubers construct and fine-tune their identities, how they perform such identities in an exaggerated way, and their unique gender expression methods. In this section, we explore our findings from three perspectives: identity co-construction, avatar-mediated streamer presentation, and feminism. Afterwards, we discuss design implications for both VTubing and live streaming in general and discuss the limitations this research and possible future work.

\subsection{Identity Co-Construction as a Form of Engagement}
This research built upon prior knowledge about identity management in avatar-mediated systems and live streaming.
Similar to other avatar-mediated systems such as social VR or virtual worlds \cite{Freeman2015,Freeman2016,Huh2010,Ducheneaut2009}, VTubing affords experimentation with virtual identities, which in turn affects the performances of Nakonohitos during live streams.
In addition, as VTubers' identities were often co-constructed between viewers and the Nakanohito, VTubers constantly fine-tuned their avatar design and performances based on viewer feedback and suggestions, or for the purpose of viewer engagement. Such co-construction constitutes a unique form of audience participation in live streams. Unlike previously studied viewer-streamer interaction such as chatting \cite{Hamilton2014,Tang2016}, polling \cite{Cheung2011,Lessel2017}, gifting \cite{Hilvert2018,Lu2018youwatch}, etc., VTubing directly enables viewers to shape the identities of streamers and how such identities should be performed. Such forms of interaction also complement the conceptualisation of co-performance found by Li et al. \cite{Li2018} since viewer participation usually translates into performances of the avatar during live streams.

The findings also suggest that such participation contributes to the engagement of VTubing. Viewers like to participate in co-construction and prefer interacting with an identity that was co-constructed rather than simply watching it play games. Therefore, VTubers tended to make their streams more interactive, however, VTubers still had the initiative and dominated the co-construction process, due to the affordances of live streaming.
Unlike avatar customization in virtual worlds or social VR,
the construction of a VTuber identity is more complicated, which grants greater agency. The high-fidelity nature of live streams empowers VTubers to endow their avatars with more complex human characters through their long-term performances and design efforts. It also requires VTubers to be more authentic \cite{Pellicone2017,Tang2022}, since performing an entirely different identity would be much more difficult than in other avatar-mediated systems such as virtual worlds. Additionally, the interactivity and synchronicity \cite{Haimson2017,Hamilton2014} of live streaming makes it possible for others to directly participate in identity construction. VTubers are often incentivized to do so to offer their viewers a more engaging experience, but they ultimately design or perform their avatars based on their preferences and identities in real life, after taking viewer feedback into account.


\subsection{Exaggeration vs. Authenticity in Self-presentation of VTubers}
The unique socio-technical arrangement of VTubing was also found to affect the self-presentation of streamers. The performance of co-constructed identities was found to become more inflated, affected, and candid. Compared to real-person streamers, avatar-mediated live streaming caused reduced non-verbal cues \cite{Guye1999,Fabri2002,Wigham2013,Maloney2020}, which often lead to the exaggeration of VTuber performances. However, such exaggerations may not have translated into selective and optimised self-presentation as suggested by the hyper-personal model \cite{Walther1996,Walther2011} and previous studies on social media \cite{Kairam2012,Xiao2020}. This is because the synchronicity, interactivity, and high fidelity of live streams forces VTubers to maintain an authentic and engaging identity. Contrarily, some VTubers turned to VTubing to present their real selves and stream what they liked to stream. This was often attributed to the pseudonymity of VTubing that facilitated self-disclosure and afforded candid and honest self-presentation, similar to other platforms supporting relative anonymity \cite{Ammari2019,Ellison2016,Leavitt2015,Ma2016,Schlesinger2017,Schoenebeck2013}. It is also worth noting that such pseudonymity would not compromise social accountability such as other social media \cite{Bernstein2011,Friedman2001}, since VTubers still needed to build a reputation based on their virtual identities. With VTubers, self-disclosure and candid self-presentation often served as methods of building intimacy and engaging viewers (e.g., C3), which complements prior work on lives streaming communities from viewer perspectives \cite{Sheng2020}.

Despite the exaggeration behaviors used,the VTubing community was generally inclusive and tolerant towards VTubers, which echos previous research on viewer perceptions of VTubers \cite{Lu2021}. Some interviewees (e.g., I2, I5) stressed the reduction in harassment and inclusiveness experienced during their VTubing practices. On the one hand, this could be attributed to the `sense of distance' proposed by Lu et al. \cite{Lu2021} that makes the Nakanohito invisible behind the scenes and reminds viewers that the inflated presentation is acting. On the other hand, the reduced non-verbal cues of VTubing may lead viewers to inflate and idealize their perception of a VTuber, as suggested by the hyper-personal model \cite{Walther1996}, which translates into a tolerant and inclusive environment.

\subsection{Revisiting VTubing through Lens of Objectification Theory}
The objectification theory posits that women are frequently sexually objectified via the male gaze \cite{Mulvey1989} and that the prevalence of such objectification can lead to internalization, also known as self-objectification, where women internalise the male gaze as the primary view of themselves \cite{Fredrickson1997,Moradi2008}. Self-objectification causes negative consequences such as body shaming, depression, anxiety, eating disorders, etc. \cite{Moradi2008,Calogero2004}. Unlike other social media or avatar-mediated systems that often put women in objectified roles (e.g., \cite{davis2018objectification}), VTubing does afford greater agency in gender-related design and performance, where VTubers, mostly women, can infuse customized gender and sexuality into their avatars via outfit design, animation design, voicing, chatting with viewers, and so on. Such affordances open avenues for future investigation into the landscapes of gender expression by disentangling virtual and real gender identities.
Particularly, we are interested in how female streamers consider their virtual gender customization before and after being a VTuber in a misogynist environment, and whether the ``female dominance'' of VTubing should be seen as an empowerment or objectification for women.

Our research focused primarily on the customization and performance of the virtual gender alone. VTubing allowed for experimentation of new gender identities (I4) and gender expressions different from stereotypes (e.g., C5's avatar design). Our interviewees also mentioned it alleviated the appearance anxiety of being a streamer and sometimes granted a sense of power (I3) and control over gender expression (I4). Nevertheless, such agency might not necessarily translate into empowered sexuality \cite{Baumgardner2004,Attwood2007,Levy2010}, especially when exposed to misogynistic viewers \cite{Lu2021}. Self-sexualization is prevalent and often used as a strategy to engage viewers (mostly male). It remains to be seen how such self-sexualization impacts psychological well-being of Nakanohitos, especially women, and whether it leads to self-objectification while VTubing via an often sexualized avatar that does not necessarily resemble themselves. We advocate for future efforts in CSCW to understand such issues through the lens of feminism. In particular, a survey study (e.g., using Twenty Statement Test \cite{fredrickson1998swimsuit}) is needed to measure potential self-objectification in this scenario. Many interviewees explained that their motivations were self-driven and such sexualization was based solely on their virtual avatar. However, this is still concerning because previous studies in lab environments suggested that controlling a sexualized avatar that resembled one's self might still lead to self-objectification in the real world \cite{Fox2013,Fox2015}.

Despite sexualized gender expression, our interviewees almost unanimously reported less sexual harassment or hostility from viewers, which is uncommon in live streaming \cite{Ruberg2019,Wotanis2014,Doring2019}. The question of whether this phenomenon can be seen as empowerment also remains open for feminist debate. VTubing could potentially act as an expedient for women or members of other minority groups to stream, however, such practices are still limited, since they make no effort to change misogynistic culture \cite{Lu2021,Ruberg2019}, but in some ways, still caters to it for viewership. Similar to the concern of `raunch culture' \cite{Levy2010}, the seeming female-dominance of VTubing that is entrenched in such a culture might actually imply a sense of false empowerment, as it works to maintain the sexual status quo.

\subsection{Design Implications}
In this subsection, we propose design implications for VTubing and beyond based on of our findings.

\subsubsection{Design for VTuber Performance}
Despite its rapid growth, VTubing is still somewhat limited in its current form. More design effort should be made to improve on the expressiveness of performances to overcome the limitations inherent in the reduction of non-verbal cues and the difficulty of coordinating avatar performances (i,e., rigging the avatar, voicing, and presenting streaming materials at the same time). For example, smart gloves or wristbands could be used to reconstruct the hand gestures of virtual avatars during live performances \cite{Ike2014,Wen2020}. To facilitate the coordination of avatar performances, there is a design opportunity to provide support tools that enable VTubers to seamlessly rig avatars and streamed content such as drawings, sketches, or shared screens, at the same time. Furthermore, as most interviewed VTubers mentioned the need for cheaper 3D avatars, there is thus a need to remove the barriers of creating a 3D avatar and support performance-driven animation of these avatars during live streams via improved software and design tools.
For example, generative AI can be used to produce affordable 3D models that can be further rigged or animated while live streaming using gesture, speech or motion inputs.

\subsubsection{Design for the Pseudonymous Identity}
VTubers were found to value both the pseudonymity an avatar offers and the authenticity of their identity and performance. As prior work highlighted the sensitivity of Nakanohitos' personal information, while calling for transparency of working conditions and fair treatment \cite{Lu2021},
we argue that the blurring line between the virtual and the real should be made more transparent and explicit to ensure the authenticity of VTubing and prevent VTubers from harassment. Further guidelines should be proposed in the VTuber community or by VTubing companies to inform new VTubers and viewers of related issues. For example, VTubers could specify `dos and don'ts' for incoming viewers in their social media groups. VTubing companies can also promote `industry consensus' to fan groups or new VTubers.
Moreover, as inspired by \cite{wu2022concerned}, we call for VTubing platforms or software to better support privacy and identity management. We anticipate features that, for instance, remind viewers during live streams what content (either the performance or identity) might be virtual or authentic, as well as what should be considered private, as configured by VTubers themselves. Such practices are expected to reduce hostility or harassment and ensure an overall enjoyable and inclusive environment.

\subsubsection{Pseudonymity for Streamers}
Many VTubers reported less harassment compared to real-person streamers, as virtual avatars could save streamers from appearance concerns and support those too shy to perform publicly. VTubing could thus empower groups that are either easily targeted by online harassment or streaming efforts that are impossible to perform using a camera feed.
For example, assistive technologies can be deployed to empower people with disabilities (PWD) to stream, including diversity features in avatar customization such as signifiers of assistive devices \cite{zhang2022s,mack2023towards} to make them feel represented, or accessibility features to support avatar performance (e.g., using gesture-driven animation \cite{jiang2023handavatar}). Moreover, the pseudonymity afforded by VTubing could also be considered for streamers to address sensitive issues or topics to facilitate self-disclosure or frank conversations with audiences. In this way the practice of VTubing might promote a more honest, frank, and authentic streaming environment.

\subsection{Limitations and Future Work}
As the VTubers involved in our qualitative study were all Chinese-speaking, cultural affordances were not taken into consideration during analysis. Moreover, although the VTubers involved were evenly distributed in affiliation status, the observations and interviews could be biased because VTubers affiliated with a company could have been less inclined to be interviewed. This means that the internal motivations of VTubers with affiliations might not be represented in the findings.
Additionally, two of our participants were using virtual identities while being interviewed. They might not fully disclose their personal self so as to maintain their identity, though agreeing to answer questions on behalf of their real selves.

We thus encourage future research to continue exploring VTubing and the use of virtual avatars while live streaming. Expressiveness in 2D avatar animation and creativity support in 3D avatar design appear to be promising fields for the foreseeable future. Furthermore, once the adoption of a virtual identity has accumulated more popularity among streamers, we expect to conduct a larger-scale study on self-presentation across cultural backgrounds to deepen our understanding of VTubers, their practises, and community.
\section{Conclusion}
This research studied how the configuration of a virtual avatar while live streaming is a reconstruction of the self-presentation of a streamer. The findings revealed that one's virtual avatar affords a blending of virtual and real identities, which in turn enacts overacting in streamers' performances and inclusiveness in viewer-streamer interactions. Live streaming practices around identity were described and the gender implications of utilizing virtual avatars were examined and discussed. This work thus provided increased knowledge about avatar-mediated live streaming and should inspire future avatar-related design and research in both avatar-mediated systems and live streaming.

\section*{Acknowledgements}
This research was partially supported by the 2021 CCF-Tencent
Rhino-Bird Research Fund and the Research Matching Grant Scheme
(RMGS, Project No. 9229095). We thank Tencent, China Computer
Federation (CCF), and Research Grants Council of Hong Kong for
their support.


\bibliographystyle{ACM-Reference-Format}
\bibliography{references}


\begin{thebibliography}{101}


\ifx \showCODEN    \undefined \def \showCODEN     #1{\unskip}     \fi
\ifx \showDOI      \undefined \def \showDOI       #1{#1}\fi
\ifx \showISBNx    \undefined \def \showISBNx     #1{\unskip}     \fi
\ifx \showISBNxiii \undefined \def \showISBNxiii  #1{\unskip}     \fi
\ifx \showISSN     \undefined \def \showISSN      #1{\unskip}     \fi
\ifx \showLCCN     \undefined \def \showLCCN      #1{\unskip}     \fi
\ifx \shownote     \undefined \def \shownote      #1{#1}          \fi
\ifx \showarticletitle \undefined \def \showarticletitle #1{#1}   \fi
\ifx \showURL      \undefined \def \showURL       {\relax}        \fi
\providecommand\bibfield[2]{#2}
\providecommand\bibinfo[2]{#2}
\providecommand\natexlab[1]{#1}
\providecommand\showeprint[2][]{arXiv:#2}

\bibitem[Ammari et~al\mbox{.}(2019)]%
        {Ammari2019}
\bibfield{author}{\bibinfo{person}{Tawfiq Ammari}, \bibinfo{person}{Sarita Schoenebeck}, {and} \bibinfo{person}{Daniel Romero}.} \bibinfo{year}{2019}\natexlab{}.
\newblock \showarticletitle{Self-declared throwaway accounts on Reddit: How platform affordances and shared norms enable parenting disclosure and support}.
\newblock \bibinfo{journal}{\emph{Proceedings of the ACM on Human-Computer Interaction}} \bibinfo{volume}{3}, \bibinfo{number}{CSCW} (\bibinfo{year}{2019}), \bibinfo{pages}{1--30}.
\newblock


\bibitem[Argyle and Dean(1965)]%
        {Argyle1965}
\bibfield{author}{\bibinfo{person}{Michael Argyle} {and} \bibinfo{person}{Janet Dean}.} \bibinfo{year}{1965}\natexlab{}.
\newblock \showarticletitle{Eye-contact, distance and affiliation}.
\newblock \bibinfo{journal}{\emph{Sociometry}} (\bibinfo{year}{1965}), \bibinfo{pages}{289--304}.
\newblock


\bibitem[Attwood(2007)]%
        {Attwood2007}
\bibfield{author}{\bibinfo{person}{Feona Attwood}.} \bibinfo{year}{2007}\natexlab{}.
\newblock \showarticletitle{Sluts and riot grrrls: Female identity and sexual agency}.
\newblock \bibinfo{journal}{\emph{Journal of gender studies}} \bibinfo{volume}{16}, \bibinfo{number}{3} (\bibinfo{year}{2007}), \bibinfo{pages}{233--247}.
\newblock


\bibitem[Baumgardner and Richards(2004)]%
        {Baumgardner2004}
\bibfield{author}{\bibinfo{person}{Jennifer Baumgardner} {and} \bibinfo{person}{Amy Richards}.} \bibinfo{year}{2004}\natexlab{}.
\newblock \showarticletitle{Feminism and Femininity: Or How We Learned to Stop Worrying}.
\newblock \bibinfo{journal}{\emph{All about the girl: Culture, power, and identity}} (\bibinfo{year}{2004}), \bibinfo{pages}{59}.
\newblock


\bibitem[Bernstein et~al\mbox{.}(2011)]%
        {Bernstein2011}
\bibfield{author}{\bibinfo{person}{Michael Bernstein}, \bibinfo{person}{Andr{\'e}s Monroy-Hern{\'a}ndez}, \bibinfo{person}{Drew Harry}, \bibinfo{person}{Paul Andr{\'e}}, \bibinfo{person}{Katrina Panovich}, {and} \bibinfo{person}{Greg Vargas}.} \bibinfo{year}{2011}\natexlab{}.
\newblock \showarticletitle{4chan and/b: An Analysis of Anonymity and Ephemerality in a Large Online Community}. In \bibinfo{booktitle}{\emph{Proceedings of the International AAAI Conference on Web and Social Media}}, Vol.~\bibinfo{volume}{5}.
\newblock


\bibitem[Bessi{\`e}re et~al\mbox{.}(2007)]%
        {Bessiere2007}
\bibfield{author}{\bibinfo{person}{Katherine Bessi{\`e}re}, \bibinfo{person}{A~Fleming Seay}, {and} \bibinfo{person}{Sara Kiesler}.} \bibinfo{year}{2007}\natexlab{}.
\newblock \showarticletitle{The ideal elf: Identity exploration in World of Warcraft}.
\newblock \bibinfo{journal}{\emph{Cyberpsychology \& behavior}} \bibinfo{volume}{10}, \bibinfo{number}{4} (\bibinfo{year}{2007}), \bibinfo{pages}{530--535}.
\newblock


\bibitem[Birk et~al\mbox{.}(2016)]%
        {Birk2016}
\bibfield{author}{\bibinfo{person}{Max~V Birk}, \bibinfo{person}{Cheralyn Atkins}, \bibinfo{person}{Jason~T Bowey}, {and} \bibinfo{person}{Regan~L Mandryk}.} \bibinfo{year}{2016}\natexlab{}.
\newblock \showarticletitle{Fostering intrinsic motivation through avatar identification in digital games}. In \bibinfo{booktitle}{\emph{Proceedings of the 2016 CHI conference on human factors in computing systems}}. \bibinfo{pages}{2982--2995}.
\newblock


\bibitem[Bopp et~al\mbox{.}(2019)]%
        {Bopp2019}
\bibfield{author}{\bibinfo{person}{Julia~Ayumi Bopp}, \bibinfo{person}{Livia~J M{\"u}ller}, \bibinfo{person}{Lena~Fanya Aeschbach}, \bibinfo{person}{Klaus Opwis}, {and} \bibinfo{person}{Elisa~D Mekler}.} \bibinfo{year}{2019}\natexlab{}.
\newblock \showarticletitle{Exploring emotional attachment to game characters}. In \bibinfo{booktitle}{\emph{Proceedings of the Annual Symposium on Computer-Human Interaction in Play}}. \bibinfo{pages}{313--324}.
\newblock


\bibitem[Cai and Wohn(2019)]%
        {Cai2019}
\bibfield{author}{\bibinfo{person}{Jie Cai} {and} \bibinfo{person}{Donghee~Yvette Wohn}.} \bibinfo{year}{2019}\natexlab{}.
\newblock \showarticletitle{Live streaming commerce: Uses and gratifications approach to understanding consumers’ motivations}. In \bibinfo{booktitle}{\emph{Proceedings of the 52nd Hawaii International Conference on System Sciences}}.
\newblock


\bibitem[Calogero(2004)]%
        {Calogero2004}
\bibfield{author}{\bibinfo{person}{Rachel~M Calogero}.} \bibinfo{year}{2004}\natexlab{}.
\newblock \showarticletitle{A test of objectification theory: The effect of the male gaze on appearance concerns in college women}.
\newblock \bibinfo{journal}{\emph{Psychology of women quarterly}} \bibinfo{volume}{28}, \bibinfo{number}{1} (\bibinfo{year}{2004}), \bibinfo{pages}{16--21}.
\newblock


\bibitem[Cheung and Huang(2011)]%
        {Cheung2011}
\bibfield{author}{\bibinfo{person}{Gifford Cheung} {and} \bibinfo{person}{Jeff Huang}.} \bibinfo{year}{2011}\natexlab{}.
\newblock \showarticletitle{Starcraft from the stands: understanding the game spectator}. In \bibinfo{booktitle}{\emph{Proceedings of the SIGCHI conference on human factors in computing systems}}. \bibinfo{pages}{763--772}.
\newblock


\bibitem[Consalvo(2017)]%
        {Consalvo2017}
\bibfield{author}{\bibinfo{person}{Mia Consalvo}.} \bibinfo{year}{2017}\natexlab{}.
\newblock \showarticletitle{Player one, playing with others virtually: What’s next in game and player studies}.
\newblock \bibinfo{journal}{\emph{Critical Studies in Media Communication}} \bibinfo{volume}{34}, \bibinfo{number}{1} (\bibinfo{year}{2017}), \bibinfo{pages}{84--87}.
\newblock


\bibitem[Corbin and Strauss(2014)]%
        {Corbin2014}
\bibfield{author}{\bibinfo{person}{Juliet Corbin} {and} \bibinfo{person}{Anselm Strauss}.} \bibinfo{year}{2014}\natexlab{}.
\newblock \bibinfo{booktitle}{\emph{Basics of qualitative research: Techniques and procedures for developing grounded theory}}.
\newblock \bibinfo{publisher}{Sage publications}.
\newblock


\bibitem[Davis(2018)]%
        {davis2018objectification}
\bibfield{author}{\bibinfo{person}{Stefanie~E Davis}.} \bibinfo{year}{2018}\natexlab{}.
\newblock \showarticletitle{Objectification, sexualization, and misrepresentation: Social media and the college experience}.
\newblock \bibinfo{journal}{\emph{Social Media+ Society}} \bibinfo{volume}{4}, \bibinfo{number}{3} (\bibinfo{year}{2018}), \bibinfo{pages}{2056305118786727}.
\newblock


\bibitem[DeVito et~al\mbox{.}(2017)]%
        {Devito2017}
\bibfield{author}{\bibinfo{person}{Michael~A DeVito}, \bibinfo{person}{Jeremy Birnholtz}, {and} \bibinfo{person}{Jeffery~T Hancock}.} \bibinfo{year}{2017}\natexlab{}.
\newblock \showarticletitle{Platforms, people, and perception: Using affordances to understand self-presentation on social media}. In \bibinfo{booktitle}{\emph{Proceedings of the 2017 ACM conference on computer supported cooperative work and social computing}}. \bibinfo{pages}{740--754}.
\newblock


\bibitem[D{\"o}ring and Mohseni(2019)]%
        {Doring2019}
\bibfield{author}{\bibinfo{person}{Nicola D{\"o}ring} {and} \bibinfo{person}{M~Rohangis Mohseni}.} \bibinfo{year}{2019}\natexlab{}.
\newblock \showarticletitle{Male dominance and sexism on YouTube: results of three content analyses}.
\newblock \bibinfo{journal}{\emph{Feminist Media Studies}} \bibinfo{volume}{19}, \bibinfo{number}{4} (\bibinfo{year}{2019}), \bibinfo{pages}{512--524}.
\newblock


\bibitem[Ducheneaut et~al\mbox{.}(2009)]%
        {Ducheneaut2009}
\bibfield{author}{\bibinfo{person}{Nicolas Ducheneaut}, \bibinfo{person}{Ming-Hui Wen}, \bibinfo{person}{Nicholas Yee}, {and} \bibinfo{person}{Greg Wadley}.} \bibinfo{year}{2009}\natexlab{}.
\newblock \showarticletitle{Body and mind: a study of avatar personalization in three virtual worlds}. In \bibinfo{booktitle}{\emph{Proceedings of the SIGCHI conference on human factors in computing systems}}. \bibinfo{pages}{1151--1160}.
\newblock


\bibitem[Ellison et~al\mbox{.}(2016)]%
        {Ellison2016}
\bibfield{author}{\bibinfo{person}{Nicole~B Ellison}, \bibinfo{person}{Lindsay Blackwell}, \bibinfo{person}{Cliff Lampe}, {and} \bibinfo{person}{Penny Trieu}.} \bibinfo{year}{2016}\natexlab{}.
\newblock \showarticletitle{“The question exists, but you don’t exist with it”: Strategic anonymity in the social lives of adolescents}.
\newblock \bibinfo{journal}{\emph{Social Media+ Society}} \bibinfo{volume}{2}, \bibinfo{number}{4} (\bibinfo{year}{2016}), \bibinfo{pages}{2056305116670673}.
\newblock


\bibitem[Faas et~al\mbox{.}(2018)]%
        {Faas2018}
\bibfield{author}{\bibinfo{person}{Travis Faas}, \bibinfo{person}{Lynn Dombrowski}, \bibinfo{person}{Alyson Young}, {and} \bibinfo{person}{Andrew~D Miller}.} \bibinfo{year}{2018}\natexlab{}.
\newblock \showarticletitle{Watch me code: Programming mentorship communities on twitch. tv}.
\newblock \bibinfo{journal}{\emph{Proceedings of the ACM on Human-Computer Interaction}} \bibinfo{volume}{2}, \bibinfo{number}{CSCW} (\bibinfo{year}{2018}), \bibinfo{pages}{1--18}.
\newblock


\bibitem[Fabri et~al\mbox{.}(2002)]%
        {Fabri2002}
\bibfield{author}{\bibinfo{person}{Marc Fabri}, \bibinfo{person}{DJ Moore}, {and} \bibinfo{person}{DJ Hobbs}.} \bibinfo{year}{2002}\natexlab{}.
\newblock \showarticletitle{Expressive agents: Non-verbal communication in collaborative virtual environments}.
\newblock \bibinfo{journal}{\emph{Proceedings of Autonomous Agents and Multi-Agent Systems (Embodied Conversational Agents)}} (\bibinfo{year}{2002}).
\newblock


\bibitem[Fox et~al\mbox{.}(2013)]%
        {Fox2013}
\bibfield{author}{\bibinfo{person}{Jesse Fox}, \bibinfo{person}{Jeremy~N Bailenson}, {and} \bibinfo{person}{Liz Tricase}.} \bibinfo{year}{2013}\natexlab{}.
\newblock \showarticletitle{The embodiment of sexualized virtual selves: The Proteus effect and experiences of self-objectification via avatars}.
\newblock \bibinfo{journal}{\emph{Computers in Human Behavior}} \bibinfo{volume}{29}, \bibinfo{number}{3} (\bibinfo{year}{2013}), \bibinfo{pages}{930--938}.
\newblock


\bibitem[Fox et~al\mbox{.}(2015)]%
        {Fox2015}
\bibfield{author}{\bibinfo{person}{Jesse Fox}, \bibinfo{person}{Rachel~A Ralston}, \bibinfo{person}{Cody~K Cooper}, {and} \bibinfo{person}{Kaitlyn~A Jones}.} \bibinfo{year}{2015}\natexlab{}.
\newblock \showarticletitle{Sexualized avatars lead to women’s self-objectification and acceptance of rape myths}.
\newblock \bibinfo{journal}{\emph{Psychology of Women Quarterly}} \bibinfo{volume}{39}, \bibinfo{number}{3} (\bibinfo{year}{2015}), \bibinfo{pages}{349--362}.
\newblock


\bibitem[Fredrickson and Roberts(1997)]%
        {Fredrickson1997}
\bibfield{author}{\bibinfo{person}{Barbara~L Fredrickson} {and} \bibinfo{person}{Tomi-Ann Roberts}.} \bibinfo{year}{1997}\natexlab{}.
\newblock \showarticletitle{Objectification theory: Toward understanding women's lived experiences and mental health risks}.
\newblock \bibinfo{journal}{\emph{Psychology of women quarterly}} \bibinfo{volume}{21}, \bibinfo{number}{2} (\bibinfo{year}{1997}), \bibinfo{pages}{173--206}.
\newblock


\bibitem[Fredrickson et~al\mbox{.}(1998)]%
        {fredrickson1998swimsuit}
\bibfield{author}{\bibinfo{person}{Barbara~L Fredrickson}, \bibinfo{person}{Tomi-Ann Roberts}, \bibinfo{person}{Stephanie~M Noll}, \bibinfo{person}{Diane~M Quinn}, {and} \bibinfo{person}{Jean~M Twenge}.} \bibinfo{year}{1998}\natexlab{}.
\newblock \showarticletitle{That swimsuit becomes you: sex differences in self-objectification, restrained eating, and math performance.}
\newblock \bibinfo{journal}{\emph{Journal of personality and social psychology}} \bibinfo{volume}{75}, \bibinfo{number}{1} (\bibinfo{year}{1998}), \bibinfo{pages}{269}.
\newblock


\bibitem[Freeman et~al\mbox{.}(2016)]%
        {Freeman2016}
\bibfield{author}{\bibinfo{person}{Guo Freeman}, \bibinfo{person}{Jeffrey Bardzell}, {and} \bibinfo{person}{Shaowen Bardzell}.} \bibinfo{year}{2016}\natexlab{}.
\newblock \showarticletitle{Revisiting computer-mediated intimacy: In-game marriage and dyadic gameplay in Audition}. In \bibinfo{booktitle}{\emph{Proceedings of the 2016 CHI Conference on Human Factors in Computing Systems}}. \bibinfo{pages}{4325--4336}.
\newblock


\bibitem[Freeman et~al\mbox{.}(2015)]%
        {Freeman2015}
\bibfield{author}{\bibinfo{person}{Guo Freeman}, \bibinfo{person}{Jeffrey Bardzell}, \bibinfo{person}{Shaowen Bardzell}, {and} \bibinfo{person}{Susan~C Herring}.} \bibinfo{year}{2015}\natexlab{}.
\newblock \showarticletitle{Simulating marriage: Gender roles and emerging intimacy in an online game}. In \bibinfo{booktitle}{\emph{Proceedings of the 18th ACM Conference on Computer Supported Cooperative Work \& Social Computing}}. \bibinfo{pages}{1191--1200}.
\newblock


\bibitem[Freeman and Maloney(2021)]%
        {Freeman2021}
\bibfield{author}{\bibinfo{person}{Guo Freeman} {and} \bibinfo{person}{Divine Maloney}.} \bibinfo{year}{2021}\natexlab{}.
\newblock \showarticletitle{Body, avatar, and me: The presentation and perception of self in social virtual reality}.
\newblock \bibinfo{journal}{\emph{Proceedings of the ACM on Human-Computer Interaction}} \bibinfo{volume}{4}, \bibinfo{number}{CSCW3} (\bibinfo{year}{2021}), \bibinfo{pages}{1--27}.
\newblock


\bibitem[Freeman and Wohn(2020)]%
        {Freeman2020}
\bibfield{author}{\bibinfo{person}{Guo Freeman} {and} \bibinfo{person}{Donghee~Yvette Wohn}.} \bibinfo{year}{2020}\natexlab{}.
\newblock \showarticletitle{Streaming your Identity: Navigating the Presentation of Gender and Sexuality through Live Streaming}.
\newblock \bibinfo{journal}{\emph{Computer Supported Cooperative Work (CSCW)}} \bibinfo{volume}{29}, \bibinfo{number}{6} (\bibinfo{year}{2020}), \bibinfo{pages}{795--825}.
\newblock


\bibitem[Friedman* and Resnick(2001)]%
        {Friedman2001}
\bibfield{author}{\bibinfo{person}{Eric~J Friedman*} {and} \bibinfo{person}{Paul Resnick}.} \bibinfo{year}{2001}\natexlab{}.
\newblock \showarticletitle{The social cost of cheap pseudonyms}.
\newblock \bibinfo{journal}{\emph{Journal of Economics \& Management Strategy}} \bibinfo{volume}{10}, \bibinfo{number}{2} (\bibinfo{year}{2001}), \bibinfo{pages}{173--199}.
\newblock


\bibitem[Goffman et~al\mbox{.}(1978)]%
        {Goffman1978}
\bibfield{author}{\bibinfo{person}{Erving Goffman} {et~al\mbox{.}}} \bibinfo{year}{1978}\natexlab{}.
\newblock \bibinfo{booktitle}{\emph{The presentation of self in everyday life}}. Vol.~\bibinfo{volume}{21}.
\newblock \bibinfo{publisher}{Harmondsworth London}.
\newblock


\bibitem[Guye-Vuill{\`e}me et~al\mbox{.}(1999)]%
        {Guye1999}
\bibfield{author}{\bibinfo{person}{Anthony Guye-Vuill{\`e}me}, \bibinfo{person}{Tolga~K Capin}, \bibinfo{person}{S Pandzic}, \bibinfo{person}{N~Magnenat Thalmann}, {and} \bibinfo{person}{Daniel Thalmann}.} \bibinfo{year}{1999}\natexlab{}.
\newblock \showarticletitle{Nonverbal communication interface for collaborative virtual environments}.
\newblock \bibinfo{journal}{\emph{Virtual Reality}} \bibinfo{volume}{4}, \bibinfo{number}{1} (\bibinfo{year}{1999}), \bibinfo{pages}{49--59}.
\newblock


\bibitem[Haimson(2018)]%
        {Haimson2018}
\bibfield{author}{\bibinfo{person}{Oliver Haimson}.} \bibinfo{year}{2018}\natexlab{}.
\newblock \showarticletitle{Social media as social transition machinery}.
\newblock \bibinfo{journal}{\emph{Proceedings of the ACM on Human-Computer Interaction}} \bibinfo{volume}{2}, \bibinfo{number}{CSCW} (\bibinfo{year}{2018}), \bibinfo{pages}{1--21}.
\newblock


\bibitem[Haimson and Tang(2017)]%
        {Haimson2017}
\bibfield{author}{\bibinfo{person}{Oliver~L Haimson} {and} \bibinfo{person}{John~C Tang}.} \bibinfo{year}{2017}\natexlab{}.
\newblock \showarticletitle{What makes live events engaging on Facebook Live, Periscope, and Snapchat}. In \bibinfo{booktitle}{\emph{Proceedings of the 2017 CHI conference on human factors in computing systems}}. \bibinfo{pages}{48--60}.
\newblock


\bibitem[Hamilton et~al\mbox{.}(2014)]%
        {Hamilton2014}
\bibfield{author}{\bibinfo{person}{William~A Hamilton}, \bibinfo{person}{Oliver Garretson}, {and} \bibinfo{person}{Andruid Kerne}.} \bibinfo{year}{2014}\natexlab{}.
\newblock \showarticletitle{Streaming on twitch: fostering participatory communities of play within live mixed media}. In \bibinfo{booktitle}{\emph{Proceedings of the SIGCHI conference on human factors in computing systems}}. \bibinfo{pages}{1315--1324}.
\newblock


\bibitem[Hamilton et~al\mbox{.}(2018)]%
        {Hamilton2018}
\bibfield{author}{\bibinfo{person}{William~A Hamilton}, \bibinfo{person}{Nic Lupfer}, \bibinfo{person}{Nicolas Botello}, \bibinfo{person}{Tyler Tesch}, \bibinfo{person}{Alex Stacy}, \bibinfo{person}{Jeremy Merrill}, \bibinfo{person}{Blake Williford}, \bibinfo{person}{Frank~R Bentley}, {and} \bibinfo{person}{Andruid Kerne}.} \bibinfo{year}{2018}\natexlab{}.
\newblock \showarticletitle{Collaborative live media curation: Shared context for participation in online learning}. In \bibinfo{booktitle}{\emph{Proceedings of the 2018 CHI Conference on Human Factors in Computing Systems}}. \bibinfo{pages}{1--14}.
\newblock


\bibitem[Hilvert-Bruce et~al\mbox{.}(2018)]%
        {Hilvert2018}
\bibfield{author}{\bibinfo{person}{Zorah Hilvert-Bruce}, \bibinfo{person}{James~T Neill}, \bibinfo{person}{Max Sj{\"o}blom}, {and} \bibinfo{person}{Juho Hamari}.} \bibinfo{year}{2018}\natexlab{}.
\newblock \showarticletitle{Social motivations of live-streaming viewer engagement on Twitch}.
\newblock \bibinfo{journal}{\emph{Computers in Human Behavior}}  \bibinfo{volume}{84} (\bibinfo{year}{2018}), \bibinfo{pages}{58--67}.
\newblock


\bibitem[Hogan(2010)]%
        {Hogan2010}
\bibfield{author}{\bibinfo{person}{Bernie Hogan}.} \bibinfo{year}{2010}\natexlab{}.
\newblock \showarticletitle{The presentation of self in the age of social media: Distinguishing performances and exhibitions online}.
\newblock \bibinfo{journal}{\emph{Bulletin of Science, Technology \& Society}} \bibinfo{volume}{30}, \bibinfo{number}{6} (\bibinfo{year}{2010}), \bibinfo{pages}{377--386}.
\newblock


\bibitem[Huh and Williams(2010)]%
        {Huh2010}
\bibfield{author}{\bibinfo{person}{Searle Huh} {and} \bibinfo{person}{Dmitri Williams}.} \bibinfo{year}{2010}\natexlab{}.
\newblock \showarticletitle{Dude looks like a lady: Gender swapping in an online game}.
\newblock In \bibinfo{booktitle}{\emph{Online worlds: Convergence of the real and the virtual}}. \bibinfo{publisher}{Springer}, \bibinfo{pages}{161--174}.
\newblock


\bibitem[Ike et~al\mbox{.}(2014)]%
        {Ike2014}
\bibfield{author}{\bibinfo{person}{Tsukasa Ike}, \bibinfo{person}{Toshiaki Nakasu}, {and} \bibinfo{person}{Yasunobu Yamauchi}.} \bibinfo{year}{2014}\natexlab{}.
\newblock \showarticletitle{Contents-aware gesture interaction using wearable motion sensor}. In \bibinfo{booktitle}{\emph{Proceedings of the 2014 ACM International Symposium on Wearable Computers: Adjunct Program}}. \bibinfo{pages}{5--8}.
\newblock


\bibitem[Ito et~al\mbox{.}(2012)]%
        {ito2012fandom}
\bibfield{author}{\bibinfo{person}{Mizuko Ito}, \bibinfo{person}{Daisuke Okabe}, {and} \bibinfo{person}{Izumi Tsuji}.} \bibinfo{year}{2012}\natexlab{}.
\newblock \bibinfo{booktitle}{\emph{Fandom unbound: Otaku culture in a connected world}}.
\newblock \bibinfo{publisher}{Yale University Press}.
\newblock


\bibitem[Jiang et~al\mbox{.}(2023)]%
        {jiang2023handavatar}
\bibfield{author}{\bibinfo{person}{Yu Jiang}, \bibinfo{person}{Zhipeng Li}, \bibinfo{person}{Mufei He}, \bibinfo{person}{David Lindlbauer}, {and} \bibinfo{person}{Yukang Yan}.} \bibinfo{year}{2023}\natexlab{}.
\newblock \showarticletitle{HandAvatar: Embodying Non-Humanoid Virtual Avatars through Hands}. In \bibinfo{booktitle}{\emph{Proceedings of the 2023 CHI Conference on Human Factors in Computing Systems}}. \bibinfo{pages}{1--17}.
\newblock


\bibitem[Kairam et~al\mbox{.}(2012)]%
        {Kairam2012}
\bibfield{author}{\bibinfo{person}{Sanjay Kairam}, \bibinfo{person}{Mike Brzozowski}, \bibinfo{person}{David Huffaker}, {and} \bibinfo{person}{Ed Chi}.} \bibinfo{year}{2012}\natexlab{}.
\newblock \showarticletitle{Talking in circles: selective sharing in google+}. In \bibinfo{booktitle}{\emph{Proceedings of the SIGCHI conference on human factors in computing systems}}. \bibinfo{pages}{1065--1074}.
\newblock


\bibitem[Kempe-Cook et~al\mbox{.}(2019)]%
        {Kempe2019}
\bibfield{author}{\bibinfo{person}{Lucas Kempe-Cook}, \bibinfo{person}{Stephen Tsung-Han Sher}, {and} \bibinfo{person}{Norman~Makoto Su}.} \bibinfo{year}{2019}\natexlab{}.
\newblock \showarticletitle{Behind the voices: The practice and challenges of esports casters}. In \bibinfo{booktitle}{\emph{Proceedings of the 2019 CHI Conference on Human Factors in Computing Systems}}. \bibinfo{pages}{1--12}.
\newblock


\bibitem[Kim et~al\mbox{.}(2012)]%
        {Kim2012}
\bibfield{author}{\bibinfo{person}{Changsoo Kim}, \bibinfo{person}{Sang-Gun Lee}, {and} \bibinfo{person}{Minchoel Kang}.} \bibinfo{year}{2012}\natexlab{}.
\newblock \showarticletitle{I became an attractive person in the virtual world: Users’ identification with virtual communities and avatars}.
\newblock \bibinfo{journal}{\emph{Computers in Human Behavior}} \bibinfo{volume}{28}, \bibinfo{number}{5} (\bibinfo{year}{2012}), \bibinfo{pages}{1663--1669}.
\newblock


\bibitem[Kolko(1999)]%
        {Kolko1999}
\bibfield{author}{\bibinfo{person}{Beth~E Kolko}.} \bibinfo{year}{1999}\natexlab{}.
\newblock \showarticletitle{Representing bodies in virtual space: The rhetoric of avatar design}.
\newblock \bibinfo{journal}{\emph{The Information Society}} \bibinfo{volume}{15}, \bibinfo{number}{3} (\bibinfo{year}{1999}), \bibinfo{pages}{177--186}.
\newblock


\bibitem[Leavitt(2015)]%
        {Leavitt2015}
\bibfield{author}{\bibinfo{person}{Alex Leavitt}.} \bibinfo{year}{2015}\natexlab{}.
\newblock \showarticletitle{" This is a Throwaway Account" Temporary Technical Identities and Perceptions of Anonymity in a Massive Online Community}. In \bibinfo{booktitle}{\emph{Proceedings of the 18th ACM conference on computer supported cooperative work \& social computing}}. \bibinfo{pages}{317--327}.
\newblock


\bibitem[Lessel et~al\mbox{.}(2017)]%
        {Lessel2017}
\bibfield{author}{\bibinfo{person}{Pascal Lessel}, \bibinfo{person}{Michael Mauderer}, \bibinfo{person}{Christian Wolff}, {and} \bibinfo{person}{Antonio Kr{\"u}ger}.} \bibinfo{year}{2017}\natexlab{}.
\newblock \showarticletitle{Let's play my way: Investigating audience influence in user-generated gaming live-streams}. In \bibinfo{booktitle}{\emph{Proceedings of the 2017 ACM International Conference on Interactive Experiences for TV and Online Video}}. \bibinfo{pages}{51--63}.
\newblock


\bibitem[Levy(2010)]%
        {Levy2010}
\bibfield{author}{\bibinfo{person}{Ariel Levy}.} \bibinfo{year}{2010}\natexlab{}.
\newblock \bibinfo{booktitle}{\emph{Female chauvinist pigs: Women and the rise of raunch culture}}.
\newblock \bibinfo{publisher}{Black Inc.}
\newblock


\bibitem[Li et~al\mbox{.}(2018)]%
        {Li2018}
\bibfield{author}{\bibinfo{person}{Yao Li}, \bibinfo{person}{Yubo Kou}, \bibinfo{person}{Je~Seok Lee}, {and} \bibinfo{person}{Alfred Kobsa}.} \bibinfo{year}{2018}\natexlab{}.
\newblock \showarticletitle{Tell me before you stream me: Managing information disclosure in video game live streaming}.
\newblock \bibinfo{journal}{\emph{Proceedings of the ACM on Human-Computer Interaction}} \bibinfo{volume}{2}, \bibinfo{number}{CSCW} (\bibinfo{year}{2018}), \bibinfo{pages}{1--18}.
\newblock


\bibitem[Lin and Wang(2014)]%
        {Lin2014}
\bibfield{author}{\bibinfo{person}{Hsin Lin} {and} \bibinfo{person}{Hua Wang}.} \bibinfo{year}{2014}\natexlab{}.
\newblock \showarticletitle{Avatar creation in virtual worlds: Behaviors and motivations}.
\newblock \bibinfo{journal}{\emph{Computers in Human Behavior}}  \bibinfo{volume}{34} (\bibinfo{year}{2014}), \bibinfo{pages}{213--218}.
\newblock


\bibitem[Litt(2012)]%
        {Litt2012}
\bibfield{author}{\bibinfo{person}{Eden Litt}.} \bibinfo{year}{2012}\natexlab{}.
\newblock \showarticletitle{Knock, knock. Who's there? The imagined audience}.
\newblock \bibinfo{journal}{\emph{Journal of broadcasting \& electronic media}} \bibinfo{volume}{56}, \bibinfo{number}{3} (\bibinfo{year}{2012}), \bibinfo{pages}{330--345}.
\newblock


\bibitem[Litt and Hargittai(2016)]%
        {Litt2016}
\bibfield{author}{\bibinfo{person}{Eden Litt} {and} \bibinfo{person}{Eszter Hargittai}.} \bibinfo{year}{2016}\natexlab{}.
\newblock \showarticletitle{The imagined audience on social network sites}.
\newblock \bibinfo{journal}{\emph{Social Media+ Society}} \bibinfo{volume}{2}, \bibinfo{number}{1} (\bibinfo{year}{2016}), \bibinfo{pages}{2056305116633482}.
\newblock


\bibitem[Lu et~al\mbox{.}(2019b)]%
        {lu2019feel}
\bibfield{author}{\bibinfo{person}{Zhicong Lu}, \bibinfo{person}{Michelle Annett}, \bibinfo{person}{Mingming Fan}, {and} \bibinfo{person}{Daniel Wigdor}.} \bibinfo{year}{2019}\natexlab{b}.
\newblock \showarticletitle{" I feel it is my responsibility to stream" Streaming and Engaging with Intangible Cultural Heritage through Livestreaming}. In \bibinfo{booktitle}{\emph{Proceedings of the 2019 CHI Conference on Human Factors in Computing Systems}}. \bibinfo{pages}{1--14}.
\newblock


\bibitem[Lu et~al\mbox{.}(2019a)]%
        {lu2019vicariously}
\bibfield{author}{\bibinfo{person}{Zhicong Lu}, \bibinfo{person}{Michelle Annett}, {and} \bibinfo{person}{Daniel Wigdor}.} \bibinfo{year}{2019}\natexlab{a}.
\newblock \showarticletitle{Vicariously experiencing it all without going outside: A study of outdoor livestreaming in China}.
\newblock \bibinfo{journal}{\emph{Proceedings of the ACM on Human-Computer Interaction}} \bibinfo{volume}{3}, \bibinfo{number}{CSCW} (\bibinfo{year}{2019}), \bibinfo{pages}{1--28}.
\newblock


\bibitem[Lu et~al\mbox{.}(2018a)]%
        {Lu2018streamwiki}
\bibfield{author}{\bibinfo{person}{Zhicong Lu}, \bibinfo{person}{Seongkook Heo}, {and} \bibinfo{person}{Daniel~J Wigdor}.} \bibinfo{year}{2018}\natexlab{a}.
\newblock \showarticletitle{Streamwiki: Enabling viewers of knowledge sharing live streams to collaboratively generate archival documentation for effective in-stream and post hoc learning}.
\newblock \bibinfo{journal}{\emph{Proceedings of the ACM on Human-Computer Interaction}} \bibinfo{volume}{2}, \bibinfo{number}{CSCW} (\bibinfo{year}{2018}), \bibinfo{pages}{1--26}.
\newblock


\bibitem[Lu et~al\mbox{.}(2021a)]%
        {lu2021streamsketch}
\bibfield{author}{\bibinfo{person}{Zhicong Lu}, \bibinfo{person}{Rubaiat~Habib Kazi}, \bibinfo{person}{Li-yi Wei}, \bibinfo{person}{Mira Dontcheva}, {and} \bibinfo{person}{Karrie Karahalios}.} \bibinfo{year}{2021}\natexlab{a}.
\newblock \showarticletitle{Streamsketch: Exploring multi-modal interactions in creative live streams}.
\newblock \bibinfo{journal}{\emph{Proceedings of the ACM on Human-Computer Interaction}} \bibinfo{volume}{5}, \bibinfo{number}{CSCW1} (\bibinfo{year}{2021}), \bibinfo{pages}{1--26}.
\newblock


\bibitem[Lu et~al\mbox{.}(2021b)]%
        {Lu2021}
\bibfield{author}{\bibinfo{person}{Zhicong Lu}, \bibinfo{person}{Chenxinran Shen}, \bibinfo{person}{Jiannan Li}, \bibinfo{person}{Hong Shen}, {and} \bibinfo{person}{Daniel Wigdor}.} \bibinfo{year}{2021}\natexlab{b}.
\newblock \showarticletitle{More Kawaii than a Real-Person Live Streamer: Understanding How the Otaku Community Engages with and Perceives Virtual YouTubers}. In \bibinfo{booktitle}{\emph{Proceedings of the 2021 CHI Conference on Human Factors in Computing Systems}}. \bibinfo{pages}{1--14}.
\newblock


\bibitem[Lu et~al\mbox{.}(2018b)]%
        {Lu2018youwatch}
\bibfield{author}{\bibinfo{person}{Zhicong Lu}, \bibinfo{person}{Haijun Xia}, \bibinfo{person}{Seongkook Heo}, {and} \bibinfo{person}{Daniel Wigdor}.} \bibinfo{year}{2018}\natexlab{b}.
\newblock \showarticletitle{You watch, you give, and you engage: a study of live streaming practices in China}. In \bibinfo{booktitle}{\emph{Proceedings of the 2018 CHI conference on human factors in computing systems}}. \bibinfo{pages}{1--13}.
\newblock


\bibitem[Ma et~al\mbox{.}(2016)]%
        {Ma2016}
\bibfield{author}{\bibinfo{person}{Xiao Ma}, \bibinfo{person}{Jeff Hancock}, {and} \bibinfo{person}{Mor Naaman}.} \bibinfo{year}{2016}\natexlab{}.
\newblock \showarticletitle{Anonymity, intimacy and self-disclosure in social media}. In \bibinfo{booktitle}{\emph{Proceedings of the 2016 CHI conference on human factors in computing systems}}. \bibinfo{pages}{3857--3869}.
\newblock


\bibitem[Mack et~al\mbox{.}(2023)]%
        {mack2023towards}
\bibfield{author}{\bibinfo{person}{Kelly Mack}, \bibinfo{person}{Rai Ching~Ling Hsu}, \bibinfo{person}{Andr{\'e}s Monroy-Hern{\'a}ndez}, \bibinfo{person}{Brian~A Smith}, {and} \bibinfo{person}{Fannie Liu}.} \bibinfo{year}{2023}\natexlab{}.
\newblock \showarticletitle{Towards Inclusive Avatars: Disability Representation in Avatar Platforms}. In \bibinfo{booktitle}{\emph{Proceedings of the 2023 CHI Conference on Human Factors in Computing Systems}}. \bibinfo{pages}{1--13}.
\newblock


\bibitem[Maloney et~al\mbox{.}(2020)]%
        {Maloney2020}
\bibfield{author}{\bibinfo{person}{Divine Maloney}, \bibinfo{person}{Guo Freeman}, {and} \bibinfo{person}{Donghee~Yvette Wohn}.} \bibinfo{year}{2020}\natexlab{}.
\newblock \showarticletitle{" Talking without a Voice" Understanding Non-verbal Communication in Social Virtual Reality}.
\newblock \bibinfo{journal}{\emph{Proceedings of the ACM on Human-Computer Interaction}} \bibinfo{volume}{4}, \bibinfo{number}{CSCW2} (\bibinfo{year}{2020}), \bibinfo{pages}{1--25}.
\newblock


\bibitem[Manninen and Kujanp{\"a}{\"a}(2007)]%
        {Manninen2007}
\bibfield{author}{\bibinfo{person}{Tony Manninen} {and} \bibinfo{person}{Tomi Kujanp{\"a}{\"a}}.} \bibinfo{year}{2007}\natexlab{}.
\newblock \showarticletitle{The value of virtual assets: the role of game characters in MMOGs}.
\newblock \bibinfo{journal}{\emph{International Journal of Business Science \& Applied Management (IJBSAM)}} \bibinfo{volume}{2}, \bibinfo{number}{1} (\bibinfo{year}{2007}), \bibinfo{pages}{21--33}.
\newblock


\bibitem[Marwick and Boyd(2011)]%
        {Marwick2011}
\bibfield{author}{\bibinfo{person}{Alice~E Marwick} {and} \bibinfo{person}{Danah Boyd}.} \bibinfo{year}{2011}\natexlab{}.
\newblock \showarticletitle{I tweet honestly, I tweet passionately: Twitter users, context collapse, and the imagined audience}.
\newblock \bibinfo{journal}{\emph{New media \& society}} \bibinfo{volume}{13}, \bibinfo{number}{1} (\bibinfo{year}{2011}), \bibinfo{pages}{114--133}.
\newblock


\bibitem[McArthur et~al\mbox{.}(2015)]%
        {Mcarthur2015}
\bibfield{author}{\bibinfo{person}{Victoria McArthur}, \bibinfo{person}{Robert~John Teather}, {and} \bibinfo{person}{Jennifer Jenson}.} \bibinfo{year}{2015}\natexlab{}.
\newblock \showarticletitle{The avatar affordances framework: mapping affordances and design trends in character creation interfaces}. In \bibinfo{booktitle}{\emph{Proceedings of the 2015 annual symposium on Computer-Human Interaction in Play}}. \bibinfo{pages}{231--240}.
\newblock


\bibitem[McDonough(1999)]%
        {Mcdonough1999}
\bibfield{author}{\bibinfo{person}{Jerome~P McDonough}.} \bibinfo{year}{1999}\natexlab{}.
\newblock \showarticletitle{Designer selves: Construction of technologically mediated identity within graphical, multiuser virtual environments}.
\newblock \bibinfo{journal}{\emph{Journal of the American Society for Information Science}} \bibinfo{volume}{50}, \bibinfo{number}{10} (\bibinfo{year}{1999}), \bibinfo{pages}{855--869}.
\newblock


\bibitem[Moradi and Huang(2008)]%
        {Moradi2008}
\bibfield{author}{\bibinfo{person}{Bonnie Moradi} {and} \bibinfo{person}{Yu-Ping Huang}.} \bibinfo{year}{2008}\natexlab{}.
\newblock \showarticletitle{Objectification theory and psychology of women: A decade of advances and future directions}.
\newblock \bibinfo{journal}{\emph{Psychology of women quarterly}} \bibinfo{volume}{32}, \bibinfo{number}{4} (\bibinfo{year}{2008}), \bibinfo{pages}{377--398}.
\newblock


\bibitem[Mulvey(1989)]%
        {Mulvey1989}
\bibfield{author}{\bibinfo{person}{Laura Mulvey}.} \bibinfo{year}{1989}\natexlab{}.
\newblock \showarticletitle{Visual pleasure and narrative cinema}.
\newblock In \bibinfo{booktitle}{\emph{Visual and other pleasures}}. \bibinfo{publisher}{Springer}, \bibinfo{pages}{14--26}.
\newblock


\bibitem[Neustaedter and Fedorovskaya({[n.\,d.]})]%
        {Neustaedter2009}
\bibfield{author}{\bibinfo{person}{Carman Neustaedter} {and} \bibinfo{person}{Elena~A Fedorovskaya}.} \bibinfo{year}{[n.\,d.]}\natexlab{}.
\newblock \showarticletitle{Presenting identity in a virtual world through avatar appearances.}
\newblock


\bibitem[Oldenburg(1999)]%
        {Oldenburg1999}
\bibfield{author}{\bibinfo{person}{Ray Oldenburg}.} \bibinfo{year}{1999}\natexlab{}.
\newblock \bibinfo{booktitle}{\emph{The great good place: Cafes, coffee shops, bookstores, bars, hair salons, and other hangouts at the heart of a community}}.
\newblock \bibinfo{publisher}{Da Capo Press}.
\newblock


\bibitem[Park and Seo(2013)]%
        {Park2013}
\bibfield{author}{\bibinfo{person}{Hyungsung Park} {and} \bibinfo{person}{Sumin Seo}.} \bibinfo{year}{2013}\natexlab{}.
\newblock \showarticletitle{Effects of collaborative activities on group identity in virtual world}.
\newblock \bibinfo{journal}{\emph{Interactive Learning Environments}} \bibinfo{volume}{21}, \bibinfo{number}{6} (\bibinfo{year}{2013}), \bibinfo{pages}{516--527}.
\newblock


\bibitem[Pellicone and Ahn(2017)]%
        {Pellicone2017}
\bibfield{author}{\bibinfo{person}{Anthony~J Pellicone} {and} \bibinfo{person}{June Ahn}.} \bibinfo{year}{2017}\natexlab{}.
\newblock \showarticletitle{The Game of Performing Play: Understanding streaming as cultural production}. In \bibinfo{booktitle}{\emph{Proceedings of the 2017 CHI conference on human factors in computing systems}}. \bibinfo{pages}{4863--4874}.
\newblock


\bibitem[Robinson(2014)]%
        {Robinson2014}
\bibfield{author}{\bibinfo{person}{Oliver~C Robinson}.} \bibinfo{year}{2014}\natexlab{}.
\newblock \showarticletitle{Sampling in interview-based qualitative research: A theoretical and practical guide}.
\newblock \bibinfo{journal}{\emph{Qualitative research in psychology}} \bibinfo{volume}{11}, \bibinfo{number}{1} (\bibinfo{year}{2014}), \bibinfo{pages}{25--41}.
\newblock


\bibitem[Ruberg et~al\mbox{.}(2019)]%
        {Ruberg2019}
\bibfield{author}{\bibinfo{person}{Bonnie Ruberg}, \bibinfo{person}{Amanda~LL Cullen}, {and} \bibinfo{person}{Kathryn Brewster}.} \bibinfo{year}{2019}\natexlab{}.
\newblock \showarticletitle{Nothing but a “titty streamer”: legitimacy, labor, and the debate over women’s breasts in video game live streaming}.
\newblock \bibinfo{journal}{\emph{Critical Studies in Media Communication}} \bibinfo{volume}{36}, \bibinfo{number}{5} (\bibinfo{year}{2019}), \bibinfo{pages}{466--481}.
\newblock


\bibitem[Ruberg and Shaw(2017)]%
        {Ruberg2017}
\bibfield{author}{\bibinfo{person}{Bonnie Ruberg} {and} \bibinfo{person}{Adrienne Shaw}.} \bibinfo{year}{2017}\natexlab{}.
\newblock \bibinfo{booktitle}{\emph{Queer game studies}}.
\newblock \bibinfo{publisher}{U of Minnesota Press}.
\newblock


\bibitem[Schlesinger et~al\mbox{.}(2017)]%
        {Schlesinger2017}
\bibfield{author}{\bibinfo{person}{Ari Schlesinger}, \bibinfo{person}{Eshwar Chandrasekharan}, \bibinfo{person}{Christina~A Masden}, \bibinfo{person}{Amy~S Bruckman}, \bibinfo{person}{W~Keith Edwards}, {and} \bibinfo{person}{Rebecca~E Grinter}.} \bibinfo{year}{2017}\natexlab{}.
\newblock \showarticletitle{Situated anonymity: Impacts of anonymity, ephemerality, and hyper-locality on social media}. In \bibinfo{booktitle}{\emph{Proceedings of the 2017 CHI conference on human factors in computing systems}}. \bibinfo{pages}{6912--6924}.
\newblock


\bibitem[Schoenebeck(2013)]%
        {Schoenebeck2013}
\bibfield{author}{\bibinfo{person}{Sarita~Yardi Schoenebeck}.} \bibinfo{year}{2013}\natexlab{}.
\newblock \showarticletitle{The secret life of online moms: Anonymity and disinhibition on youbemom. com}. In \bibinfo{booktitle}{\emph{Seventh International AAAI Conference on Weblogs and Social Media}}.
\newblock


\bibitem[Schroeder(2012)]%
        {Schroeder2012}
\bibfield{author}{\bibinfo{person}{Ralph Schroeder}.} \bibinfo{year}{2012}\natexlab{}.
\newblock \bibinfo{booktitle}{\emph{The social life of avatars: Presence and interaction in shared virtual environments}}.
\newblock \bibinfo{publisher}{Springer Science \& Business Media}.
\newblock


\bibitem[Seering et~al\mbox{.}(2017)]%
        {Seering2017}
\bibfield{author}{\bibinfo{person}{Joseph Seering}, \bibinfo{person}{Robert Kraut}, {and} \bibinfo{person}{Laura Dabbish}.} \bibinfo{year}{2017}\natexlab{}.
\newblock \showarticletitle{Shaping pro and anti-social behavior on twitch through moderation and example-setting}. In \bibinfo{booktitle}{\emph{Proceedings of the 2017 ACM conference on computer supported cooperative work and social computing}}. \bibinfo{pages}{111--125}.
\newblock


\bibitem[Sheng and Kairam(2020)]%
        {Sheng2020}
\bibfield{author}{\bibinfo{person}{Jeff~T Sheng} {and} \bibinfo{person}{Sanjay~R Kairam}.} \bibinfo{year}{2020}\natexlab{}.
\newblock \showarticletitle{From Virtual Strangers to IRL Friends: Relationship Development in Livestreaming Communities on Twitch}.
\newblock \bibinfo{journal}{\emph{Proceedings of the ACM on Human-Computer Interaction}} \bibinfo{volume}{4}, \bibinfo{number}{CSCW2} (\bibinfo{year}{2020}), \bibinfo{pages}{1--34}.
\newblock


\bibitem[Stets and Burke(2000)]%
        {Stets2000}
\bibfield{author}{\bibinfo{person}{Jan~E Stets} {and} \bibinfo{person}{Peter~J Burke}.} \bibinfo{year}{2000}\natexlab{}.
\newblock \showarticletitle{Identity theory and social identity theory}.
\newblock \bibinfo{journal}{\emph{Social psychology quarterly}} (\bibinfo{year}{2000}), \bibinfo{pages}{224--237}.
\newblock


\bibitem[Tang et~al\mbox{.}(2016)]%
        {Tang2016}
\bibfield{author}{\bibinfo{person}{John~C Tang}, \bibinfo{person}{Gina Venolia}, {and} \bibinfo{person}{Kori~M Inkpen}.} \bibinfo{year}{2016}\natexlab{}.
\newblock \showarticletitle{Meerkat and periscope: I stream, you stream, apps stream for live streams}. In \bibinfo{booktitle}{\emph{Proceedings of the 2016 CHI conference on human factors in computing systems}}. \bibinfo{pages}{4770--4780}.
\newblock


\bibitem[Tang et~al\mbox{.}(2022)]%
        {Tang2022}
\bibfield{author}{\bibinfo{person}{Ningjing Tang}, \bibinfo{person}{Lei Tao}, \bibinfo{person}{Bo Wen}, {and} \bibinfo{person}{Zhicong Lu}.} \bibinfo{year}{2022}\natexlab{}.
\newblock \showarticletitle{Dare to Dream, Dare to Livestream: How E-Commerce Livestreaming Empowers Chinese Rural Women}. In \bibinfo{booktitle}{\emph{CHI Conference on Human Factors in Computing Systems}}. \bibinfo{pages}{1--13}.
\newblock


\bibitem[Taylor(2018)]%
        {Taylor2018}
\bibfield{author}{\bibinfo{person}{TL Taylor}.} \bibinfo{year}{2018}\natexlab{}.
\newblock \bibinfo{booktitle}{\emph{Watch me play: Twitch and the rise of game live streaming}}.
\newblock \bibinfo{publisher}{Princeton University Press}.
\newblock


\bibitem[Van~Looy et~al\mbox{.}(2012)]%
        {Van2012}
\bibfield{author}{\bibinfo{person}{Jan Van~Looy}, \bibinfo{person}{C{\'e}dric Courtois}, \bibinfo{person}{Melanie De~Vocht}, {and} \bibinfo{person}{Lieven De~Marez}.} \bibinfo{year}{2012}\natexlab{}.
\newblock \showarticletitle{Player identification in online games: Validation of a scale for measuring identification in MMOGs}.
\newblock \bibinfo{journal}{\emph{Media Psychology}} \bibinfo{volume}{15}, \bibinfo{number}{2} (\bibinfo{year}{2012}), \bibinfo{pages}{197--221}.
\newblock


\bibitem[Walther(1996)]%
        {Walther1996}
\bibfield{author}{\bibinfo{person}{Joseph~B Walther}.} \bibinfo{year}{1996}\natexlab{}.
\newblock \showarticletitle{Computer-mediated communication: Impersonal, interpersonal, and hyperpersonal interaction}.
\newblock \bibinfo{journal}{\emph{Communication research}} \bibinfo{volume}{23}, \bibinfo{number}{1} (\bibinfo{year}{1996}), \bibinfo{pages}{3--43}.
\newblock


\bibitem[Walther(2011)]%
        {Walther2011}
\bibfield{author}{\bibinfo{person}{Joseph~B Walther}.} \bibinfo{year}{2011}\natexlab{}.
\newblock \showarticletitle{Theories of computer-mediated communication and interpersonal relations}.
\newblock \bibinfo{journal}{\emph{The handbook of interpersonal communication}}  \bibinfo{volume}{4} (\bibinfo{year}{2011}), \bibinfo{pages}{443--479}.
\newblock


\bibitem[Wen et~al\mbox{.}(2020)]%
        {Wen2020}
\bibfield{author}{\bibinfo{person}{Feng Wen}, \bibinfo{person}{Zhongda Sun}, \bibinfo{person}{Tianyiyi He}, \bibinfo{person}{Qiongfeng Shi}, \bibinfo{person}{Minglu Zhu}, \bibinfo{person}{Zixuan Zhang}, \bibinfo{person}{Lianhui Li}, \bibinfo{person}{Ting Zhang}, {and} \bibinfo{person}{Chengkuo Lee}.} \bibinfo{year}{2020}\natexlab{}.
\newblock \showarticletitle{Machine learning glove using self-powered conductive superhydrophobic triboelectric textile for gesture recognition in VR/AR applications}.
\newblock \bibinfo{journal}{\emph{Advanced science}} \bibinfo{volume}{7}, \bibinfo{number}{14} (\bibinfo{year}{2020}), \bibinfo{pages}{2000261}.
\newblock


\bibitem[Wigham and Chanier(2013)]%
        {Wigham2013}
\bibfield{author}{\bibinfo{person}{Ciara~R Wigham} {and} \bibinfo{person}{Thierry Chanier}.} \bibinfo{year}{2013}\natexlab{}.
\newblock \showarticletitle{A study of verbal and nonverbal communication in Second Life--the ARCHI21 experience}.
\newblock \bibinfo{journal}{\emph{ReCALL}} \bibinfo{volume}{25}, \bibinfo{number}{1} (\bibinfo{year}{2013}), \bibinfo{pages}{63--84}.
\newblock


\bibitem[Wohn et~al\mbox{.}(2018)]%
        {Wohn2018}
\bibfield{author}{\bibinfo{person}{Donghee~Yvette Wohn}, \bibinfo{person}{Guo Freeman}, {and} \bibinfo{person}{Caitlin McLaughlin}.} \bibinfo{year}{2018}\natexlab{}.
\newblock \showarticletitle{Explaining viewers' emotional, instrumental, and financial support provision for live streamers}. In \bibinfo{booktitle}{\emph{Proceedings of the 2018 CHI conference on human factors in computing systems}}. \bibinfo{pages}{1--13}.
\newblock


\bibitem[Wolfendale(2007)]%
        {Wolfendale2007}
\bibfield{author}{\bibinfo{person}{Jessica Wolfendale}.} \bibinfo{year}{2007}\natexlab{}.
\newblock \showarticletitle{My avatar, my self: Virtual harm and attachment}.
\newblock \bibinfo{journal}{\emph{Ethics and information technology}} \bibinfo{volume}{9}, \bibinfo{number}{2} (\bibinfo{year}{2007}), \bibinfo{pages}{111--119}.
\newblock


\bibitem[Wotanis and McMillan(2014)]%
        {Wotanis2014}
\bibfield{author}{\bibinfo{person}{Lindsey Wotanis} {and} \bibinfo{person}{Laurie McMillan}.} \bibinfo{year}{2014}\natexlab{}.
\newblock \showarticletitle{Performing gender on YouTube: How Jenna Marbles negotiates a hostile online environment}.
\newblock \bibinfo{journal}{\emph{Feminist Media Studies}} \bibinfo{volume}{14}, \bibinfo{number}{6} (\bibinfo{year}{2014}), \bibinfo{pages}{912--928}.
\newblock


\bibitem[Wu et~al\mbox{.}(2022)]%
        {wu2022concerned}
\bibfield{author}{\bibinfo{person}{Yanlai Wu}, \bibinfo{person}{Yao Li}, {and} \bibinfo{person}{Xinning Gui}.} \bibinfo{year}{2022}\natexlab{}.
\newblock \showarticletitle{``I Am Concerned, But...'': Streamers' Privacy Concerns and Strategies In Live Streaming Information Disclosure}.
\newblock \bibinfo{journal}{\emph{Proceedings of the ACM on Human-Computer Interaction}} \bibinfo{volume}{6}, \bibinfo{number}{CSCW2} (\bibinfo{year}{2022}), \bibinfo{pages}{1--31}.
\newblock


\bibitem[Wulf et~al\mbox{.}(2020)]%
        {Wulf2020}
\bibfield{author}{\bibinfo{person}{Tim Wulf}, \bibinfo{person}{Frank~M Schneider}, {and} \bibinfo{person}{Stefan Beckert}.} \bibinfo{year}{2020}\natexlab{}.
\newblock \showarticletitle{Watching players: An exploration of media enjoyment on Twitch}.
\newblock \bibinfo{journal}{\emph{Games and culture}} \bibinfo{volume}{15}, \bibinfo{number}{3} (\bibinfo{year}{2020}), \bibinfo{pages}{328--346}.
\newblock


\bibitem[Xiao et~al\mbox{.}(2020)]%
        {Xiao2020}
\bibfield{author}{\bibinfo{person}{Sijia Xiao}, \bibinfo{person}{Dana{\"e} Metaxa}, \bibinfo{person}{Joon~Sung Park}, \bibinfo{person}{Karrie Karahalios}, {and} \bibinfo{person}{Niloufar Salehi}.} \bibinfo{year}{2020}\natexlab{}.
\newblock \showarticletitle{Random, messy, funny, raw: finstas as intimate reconfigurations of social media}. In \bibinfo{booktitle}{\emph{Proceedings of the 2020 CHI Conference on Human Factors in Computing Systems}}. \bibinfo{pages}{1--13}.
\newblock


\bibitem[Yee and Bailenson(2007)]%
        {Yee2007Proteus}
\bibfield{author}{\bibinfo{person}{Nick Yee} {and} \bibinfo{person}{Jeremy Bailenson}.} \bibinfo{year}{2007}\natexlab{}.
\newblock \showarticletitle{The Proteus effect: The effect of transformed self-representation on behavior}.
\newblock \bibinfo{journal}{\emph{Human communication research}} \bibinfo{volume}{33}, \bibinfo{number}{3} (\bibinfo{year}{2007}), \bibinfo{pages}{271--290}.
\newblock


\bibitem[Yee et~al\mbox{.}(2009)]%
        {Yee2009}
\bibfield{author}{\bibinfo{person}{Nick Yee}, \bibinfo{person}{Jeremy~N Bailenson}, {and} \bibinfo{person}{Nicolas Ducheneaut}.} \bibinfo{year}{2009}\natexlab{}.
\newblock \showarticletitle{The Proteus effect: Implications of transformed digital self-representation on online and offline behavior}.
\newblock \bibinfo{journal}{\emph{Communication Research}} \bibinfo{volume}{36}, \bibinfo{number}{2} (\bibinfo{year}{2009}), \bibinfo{pages}{285--312}.
\newblock


\bibitem[Yee et~al\mbox{.}(2007)]%
        {Yee2007Intimacy}
\bibfield{author}{\bibinfo{person}{Nick Yee}, \bibinfo{person}{Jeremy~N Bailenson}, \bibinfo{person}{Mark Urbanek}, \bibinfo{person}{Francis Chang}, {and} \bibinfo{person}{Dan Merget}.} \bibinfo{year}{2007}\natexlab{}.
\newblock \showarticletitle{The unbearable likeness of being digital: The persistence of nonverbal social norms in online virtual environments}.
\newblock \bibinfo{journal}{\emph{CyberPsychology \& Behavior}} \bibinfo{volume}{10}, \bibinfo{number}{1} (\bibinfo{year}{2007}), \bibinfo{pages}{115--121}.
\newblock


\bibitem[Yee et~al\mbox{.}(2011)]%
        {Yee2011}
\bibfield{author}{\bibinfo{person}{Nick Yee}, \bibinfo{person}{Nicolas Ducheneaut}, \bibinfo{person}{Mike Yao}, {and} \bibinfo{person}{Les Nelson}.} \bibinfo{year}{2011}\natexlab{}.
\newblock \showarticletitle{Do men heal more when in drag? Conflicting identity cues between user and avatar}. In \bibinfo{booktitle}{\emph{Proceedings of the SIGCHI conference on Human factors in computing systems}}. \bibinfo{pages}{773--776}.
\newblock


\bibitem[Yu and Young(2008)]%
        {Yu2008}
\bibfield{author}{\bibinfo{person}{Chia-Ping Yu} {and} \bibinfo{person}{Mei-Lein Young}.} \bibinfo{year}{2008}\natexlab{}.
\newblock \showarticletitle{The virtual group identification process: A virtual educational community case}.
\newblock \bibinfo{journal}{\emph{CyberPsychology \& Behavior}} \bibinfo{volume}{11}, \bibinfo{number}{1} (\bibinfo{year}{2008}), \bibinfo{pages}{87--90}.
\newblock


\bibitem[Zhang et~al\mbox{.}(2022)]%
        {zhang2022s}
\bibfield{author}{\bibinfo{person}{Kexin Zhang}, \bibinfo{person}{Elmira Deldari}, \bibinfo{person}{Zhicong Lu}, \bibinfo{person}{Yaxing Yao}, {and} \bibinfo{person}{Yuhang Zhao}.} \bibinfo{year}{2022}\natexlab{}.
\newblock \showarticletitle{“It’s Just Part of Me:” Understanding Avatar Diversity and Self-presentation of People with Disabilities in Social Virtual Reality}. In \bibinfo{booktitle}{\emph{Proceedings of the 24th International ACM SIGACCESS Conference on Computers and Accessibility}}. \bibinfo{pages}{1--16}.
\newblock


\bibitem[Zhao et~al\mbox{.}(2013)]%
        {Zhao2013}
\bibfield{author}{\bibinfo{person}{Xuan Zhao}, \bibinfo{person}{Niloufar Salehi}, \bibinfo{person}{Sasha Naranjit}, \bibinfo{person}{Sara Alwaalan}, \bibinfo{person}{Stephen Voida}, {and} \bibinfo{person}{Dan Cosley}.} \bibinfo{year}{2013}\natexlab{}.
\newblock \showarticletitle{The many faces of Facebook: Experiencing social media as performance, exhibition, and personal archive}. In \bibinfo{booktitle}{\emph{Proceedings of the SIGCHI conference on human factors in computing systems}}. \bibinfo{pages}{1--10}.
\newblock


\end{thebibliography}


\end{document}